\documentstyle[aaspp4]{article}
\def\plotonesc#1#2{\begin{center} \leavevmode
\epsfxsize=#2\columnwidth \epsfbox{#1} \end{center}}
\def\unsetyr{\def\oyear{\relax}\def\cyear{\relax}\def\cyeara{a\relax}\def\cyearb{b\relax}}
\def\setyr{\def\oyear{(}\def\cyear{)}\def\cyeara{a)}\def\cyearb{b)}}
\unsetyr
\def\jcite#1{\setyr\cite{#1}\unsetyr}
\def\rmmat#1{{\hbox{\rm #1}}}
\def\rmscr#1{\rmmat{\scriptsize #1}}
\newcommand{\be}{\begin{equation}}
\newcommand{\ee}{\end{equation}}
\newcommand{\ba}{\begin{eqnarray}}
\newcommand{\ea}{\end{eqnarray}}
\newcommand{\ie}{{\it i.e.~}}
\newcommand{\eg}{{\it e.g.~}}
\newcommand{\bt}{\begin{table} \begin{center}}
\newcommand{\et}{\end{center} \end{table}}
\def\p{\partial}
\def\d{{\rm d}}

\def\dd#1#2{\frac{\d #1}{\d #2}}
\def\pp#1#2{\frac{\p #1}{\p #2}}
\newcommand{\comment}[1]{\relax}
\def\eqref#1{Equation~\ref{eq:#1}}
\def\figref#1{Figure~\ref{fig:#1}}
\def\tabref#1{Table~\ref{tab:#1}}

\newcommand{\figcom}[1]{#1}

\newcommand{\me}{m_e}
\newcommand{\mcs}{{m_e c^2}}

\newcommand{\inoi}{\int_1^\infty}

\newcommand{\kzz}{\kappa_{zz}}
\newcommand{\kyy}{\kappa_{yy}}
\newcommand{\qos}{{q_0^2}}
\newcommand{\wpad}{{w + a_d}}

\begin{document}
\title{Multidimensional Thermal Structure of Magnetized
Neutron Star Envelopes}
\author{Jeremy S. Heyl\altaffilmark{1}}
\authoremail{jsheyl@tapir.caltech.edu}
\author{Lars Hernquist}
\authoremail{lars@ucolick.org}
\affil{Lick Observatory,
University of California, Santa Cruz, California 95064, USA}
\altaffiltext{1}{Current address: Theoretical Astrophysics, mail code 130-33,
California Institute of Technology, Pasadena, CA 91125}

\begin{abstract}

Recently launched x-ray telescopes have discovered several candidate
isolated neutron stars.  The thermal radiation from these objects may
potentially constrain our understanding of nuclear physics in a realm
inaccessible to terrestrial experiments.  To translate the observed
fluxes from neutron stars into constraints, one needs precise
calculations of the heat transfer through the thin insulating
envelopes of neutron stars.  We describe models of the thermal
structure of the envelopes of neutron stars with magnetic fields up to
$10^{14}$~G.  Unlike earlier work, we infer the properties of envelope
models in two dimensions and precisely account for the quantization of
the electron phase space.  Both dipole and uniformly magnetized
envelopes are considered.

\end{abstract}

\section{Introduction}

Isolated neutron stars can be used as powerful tools for understanding
the properties of nuclear matter.  The internal structure of neutron
stars spans the high-density, low-temperature regime of the QCD phase
diagram which is neither constrained by terrestrial experiments such
as heavy-ion collisions nor by the properties of the early universe.
Both the sizes of neutron stars and their cooling evolution depend
crucially on the nuclear equation of state and the species present in
the stellar core.  The emission that we observe from the surfaces of
isolated neutron stars does give a picture of the properties of the
nuclear material in the core, but it is a view through the crust of
the star.  Depending on the properties of the crust, the thermal flux
can vary significantly for a fixed core temperature.

Fortunately, after the first few hundred years of a neutron star's
life, the core becomes nearly isothermal and cooling proceeds
quasistatically.  During this era, the neutron star separates
thermally into three regions.  At the highest densities is the core
which provides the thermal inertia.  From densities $\rho \sim
10^{10}$~g/cm$^3$ down to $\rho \sim 10^2$~g/cm$^3$ is the envelope
which insulates the core thermally from the exterior and which
throttles the photon flux from the surface.  The lowest density region
is the atmosphere which effectively determines the spectrum of the
neutron star but not the total flux emitted.  In this paper, we focus
on the properties of the envelope which will determine the gross
properties of the emitted thermal radiation.

Although analytic studies (\eg \cite{Heyl97analns}; hereafter Paper~I)
of neutron star envelopes can well characterize the emission from
cooling neutron stars, particularly those with either sufficiently
weak or strong fields, most potentially observable neutron stars
possess field strengths in neither of these limits and have high core
temperatures which invalidate the low-temperature approximation used
in Paper~I.

The analytic technique outlined in Paper~I assumes that only the first
Landau level is filled, that the transition from the highly
non-degenerate regime to the highly degenerate regime is abrupt, and
that locally either electrons or photons dominate the heat transfer.
In the calculations here, all three of these assumptions are relaxed,
and the equations of thermal structure are integrated using opacities
that approach those used in Paper~I in the low and high temperature
limits (\cite{Pavl77}; \cite{Sila80}; \cite{Hern84a}).

In Paper~I several two dimensional models of neutron-star envelopes
were presented.  Unfortunately, the analytic separation of the
structure equation requires that the two-dimensional models be
restricted to cases where the entire degenerate portion of the
envelope is the liquid state.  Here, because the equations are solved
numerically, this restriction is not important.  We construct several
uniformly magnetized envelopes whose thermal flux is proportional to
the square of the cosine of the angle between the magnetic field and
the normal, the $\cos^2 \psi$ rule proposed earlier, and verify that
this distribution yields a uniform core temperature more generally.
These two-dimensional results are compared with those of
\jcite{Scha90a} and extended to include envelopes with a dipole field
structure.

Although it would be straightforward to examine additional effects
such as Coulomb corrections (\eg \cite{VanR88,Thor97}), for clarity
and brevity only the processes included in Paper~I are incorporated
and the consequences of our approximations are determined.

\section{The Physical Description of the Envelope}

Because the assumptions made in Paper~I will be relaxed, it is useful
to summarize the complete set of equations governing the thermal
structure of neutron star envelopes.  Again, as argued earlier, a
plane-parallel treatment is suitable for the problem.  In what
follows, use will be made of the dimensionless units
\ba
\beta &=&  \frac{\hbar \omega_B}{\mcs} = \frac{\hbar |e|}{\me^2 c^3} B
\approx \frac{B}{4.4 \times 10^{13} \rmmat { G}}, \\
\tau &=& \frac{k T}{\mcs} \approx \frac{T}{5.9 \times 10^9 \rmmat{ K}}, \\
\gamma &=& \frac{E}{\mcs} \rmmat{~and~} \zeta = \frac{\mu}{\mcs} 
\ea
where $E$ is the energy of an electron and $\mu$ is the chemical
potential of the electron gas.  It is also convenient to define $\eta
= (\zeta-1)/\tau$.

\subsection{The Thermal Structure Equation}

If we assume that the pressure is supplied by the electrons alone, the
general relativistic equations of thermal structure in the
plane-parallel approximation assume the simple form (\cite{Hern85})
\ba
\dd{\tau}{\rho} &=& \left [ \left . \pp{\rho}{\tau} \right |_\zeta + \left (
\frac{m_u}{Y_e} \frac{\kappa}{F/g_s} - \frac{S_e}{n_e} \right )
\frac{1}{k} \left . \pp{\rho}{\zeta} \right |_\tau \right ]^{-1} 
\label{eq:dtaudrho}
\\
\dd{\zeta}{\rho} &=& \left ( \left . \pp{\rho}{\zeta} \right |_\tau \right
)^{-1} \left ( 1 - \left . \pp{\rho}{\tau} \right |_\zeta
\dd{\tau}{\rho} \right ) 
\label{eq:dzetadrho} \\
\dd{z}{\rho} &=& -\dd{\zeta}{\rho} \frac{\mcs}{g_s} 
\left ( 1 - \frac{F}{g_s} \frac{S_e}{n_e \kappa} \right )^{-1}
\label{eq:dzdrho}
\ea
where $Y_e=Z/A$, $m_u$ is the atomic mass unit, $F$ is the flux
transmitted through the envelope, $g_s$ the acceleration of gravity as
measured at the surface, and $S_e$ and $n_e$ are the entropy and
number density of the electron gas.  Here, $Z$ and $A$ are the mean
atomic number and mean atomic mass of the material.  For partially
ionized matter, $Y_e$ is given by the product of $Z/A$ and the ionized
fraction.

We will further assume that the magnetic field is locally uniform and
inclined relative to the vertical by the constant angle $\psi$.  This
approximation is valid provided that the field does not vary in
direction or magnitude over the scale of the thickness of the envelope
($h_E$), \ie $|B/\nabla B| \gg h_E$.  For a multipole of order $n$,
this is equivalent to $R/n \gg h_E$, where $R$ is the stellar radius,
which holds for $n \ll 100$.  In this case,
Equations~\ref{eq:dtaudrho} through \ref{eq:dzdrho} remain valid, but
the thermal conductivity, $\kappa$, is now the sum of two
contributions \be \kappa = \kzz \cos^2 \psi + \kyy \sin^2 \psi \ee
where the field is taken to point in the $z$ direction.  Provided that
the crust is thin and the field does not vary significantly through
the envelope, it can be shown that the multidimensional equations of
thermal structure reduce to Equations~\ref{eq:dtaudrho} through
\ref{eq:dzdrho} with $\kappa$ given by the above relation (\eg Paper
I).

\subsection{Thermodynamics, Equation of State and Conductivities}

In an extremely strong magnetic field, the quantization of electron
energies into Landau levels restricts the phase space of the otherwise
free electron gas.  \jcite{Hern85} describes how to calculate the
thermodynamic quantities necessary for integrating the structure of
the neutron-star envelope.   Further details of the thermodynamic
calculations are also available in \jcite{Heyl98thesis}.

Throughout the envelope, the heat is carried by electrons and photons.
In the degenerate regime, electrons dominate the heat transfer and in
the non-degenerate regime photons carry most of the heat.

\subsubsection{Photon Conduction}

As we argued in Paper~I, for the envelopes that we will examine,
photon conduction is impeded mainly by free-free interactions with the
electron gas rather than by electron scattering.  Regardless,
corrections to the free-free opacity have a negligible influence on
the flux-core temperature relation (\cite{Hern84b}).  Here, we will
use the thermal conductivities tabulated by \jcite{Sila80} and the
analytic expressions of \jcite{Pavl77}.

\subsubsection{Electron conduction}

For the electron conductivities, we use the calculations of
\jcite{Hern84a}.  Although the analytic formulae for the parallel
conductivity derived by \jcite{Pote96b} are convenient, corresponding
results for the transverse heat flow are lacking; therefore, for
consistency, we use the parallel and perpendicular thermal
conductivities of \jcite{Hern84a}.

\paragraph{The perturbations to the distribution function.}

The results of \jcite{Hern84a} take the form of perturbations to
the distribution function ($\phi$ and $Q$) induced by a temperature
gradient for various scattering processes.
The functions $\phi$ and $Q$ are laborious to calculate.  \jcite{Hern84a}
gives fitting formulae for $n<30$ at field strengths of $B=10^{11}$,
$10^{11.5}$, $10^{12}$, $10^{12.5}$, $10^{13}$, $10^{13.5}$ and
$10^{14}$~G.  Additionally, to calculate the thermal structure of an
envelope with an dipolar magnetic field, we have calculated the function
$\phi$ at $B=2 \times 10^{12}$ and $2 \times 10^{13}$~G for $n<30$ and
at $B=2 \times 10^{14}$~G for $n<35$.  Therefore, for fields $B\geq 2
\times 10^{14}$~G, the effect of the quantization of the electron phase
space on the conductivity is included through the entire envelope. 

When only one Landau level is filled, the functions $\phi$ and $Q$ may
be expressed analytically.  In this limit, we use the following
expressions
\ba
\phi_{ep}(\gamma;\beta) &=& \frac{1}{8} w \left [ e^w E_1(w) \right ]^{-1},
\label{eq:phieplow2} \\
Q_{ep}(\gamma;\beta) &=& \beta + \frac{4}{w} - 2 e^w E_1(w) \\
\label{eq:qeplow2}
\phi_{ei} (\gamma;\beta) &=& \frac{1}{8} w
\left [
\frac{1}{\wpad} - \exp ( \wpad ) E_1 ( \wpad )
\right ]^{-1},
\label{eq:phieilow2} \\
Q_{ei} (\gamma;\beta) &=& \frac{1}{\qos} \biggr \{
\left [ \beta (a_d + 1) + 2 \qos \right ]
\exp ( \wpad ) E_1 ( \wpad  )
\nonumber \\
& & 
+ \beta (a_d + 1) (\qos +1) e^{a_d} E_1 (a_d) - \beta (2+\qos) \biggr \}
\label{eq:qeilow2}
\ea
where
\ba
\qos &=& \gamma^2 - 1,
\label{eq:qosdef2} \\
w &=& \frac{2 \qos}{\beta},
\label{eq:wdef2} \\
a_d &=& 0.15 \left ( \frac{\qos}{2 \beta} \right )^{1/3},
\label{eq:addef2}
\ea
and $E_n(x)$ is an exponential integral which is easily calculated and is
defined by
\be
E_n(x) = \inoi \frac{e^{-xt}}{t^n} dt, ~~~ x \ge 0, ~~n=0,1,\ldots
\label{eq:endef2}
\ee

\paragraph{Unquantized limit.}
To extend the conductivities beyond the maximum Landau level tabulated
we use equations (186) through (189) of \jcite{Hern84a} to calculate 
the unmagnetized counterparts of $\phi$ and $Q$.  In the liquid state
we obtain,
\ba
\phi_{ei} ( \gamma; \beta) &=& \frac{\left ( \gamma^2 - 1 \right )^3}{6 \beta^3
\zeta^2 \Lambda_{ei}}, \\
Q_{ei} ( \gamma; \beta) &=& \frac{8 \Lambda_{ei} \gamma^2}{3},
\ea
where the Coulomb logarithm $\Lambda_{ei}$ is set to ensure continuity
between the unquantized limit and the quantized calculations.  It ranges from 1 at $10^{11}$~G to
0.55 at $10^{14}$~G.  In the solid state we find
\ba
\phi_{ep} ( \gamma; \beta) &=& \frac{\left ( \gamma^2 - 1 \right )^2}{3 \beta^2
\left ( \gamma^2 + 1 \right )}, \\
Q_{ep} ( \gamma; \beta) &=& \frac{4 \left ( \gamma^4 - 1 \right)}{3 \beta}
\ea

\subsection{Numerical Integration of the Envelope}

To determine the thermal structure of the neutron star envelope, we
integrate Equations~\ref{eq:dtaudrho} through~\ref{eq:dzdrho} using the
photospheric boundary condition (\eg \cite{Kipp90})
\be
P_e = \frac{2}{3} \frac{g_s}{\tilde \kappa} = \frac{g_s \kappa \rho}{8 \sigma T^3}
\ee
where $\tilde \kappa$ is the opacity and $\sigma$ is the
Stefan-Boltzmann constant.  We omit \eqref{dzetadrho} from the system
opting instead to solve for $\zeta$ by inverting
$n_e(\zeta,\tau;\beta)$.  The system is integrated with $\ln \rho$ as
the independent variable using a Runge-Kutta method with adaptive step
size control (\cite{Pres88}).  The properties of the envelope are
calculated at 200 equally spaced steps in $\ln \rho$.
A smaller stepsize results in an unacceptable accumulation of 
roundoff errors.  The envelopes are integrated up to a density of 
$10^{10}$~g~cm$^{-3}$. 

\section{Results}

In this section we present the results of these numerical calculations.
Specifically, we focus on several aspects of the envelopes: the thermal
structure itself, the effect of dipolar fields on the moment of
inertia of the envelope, the angular dependence of the flux for a
constant field strength and the relationship between the transmitted
flux and core temperature.

\subsection{Thermal Structure}

\paragraph{Parallel Conduction.}

If the quantization of the electron phase space is neglected, the
magnetic field has no effect on the thermal conduction in the
degenerate regime.  We find that because this quantization cannot be
neglected for $B>10^{12}$~G, the magnetic field modifies the flux-core
temperature relation, especially for relatively cool neutron stars.
The case of parallel conduction has been treated in detail by
\jcite{Hern85} and \jcite{VanR88}, and we find similar results here.

\figcom{\begin{figure}
\plottwo{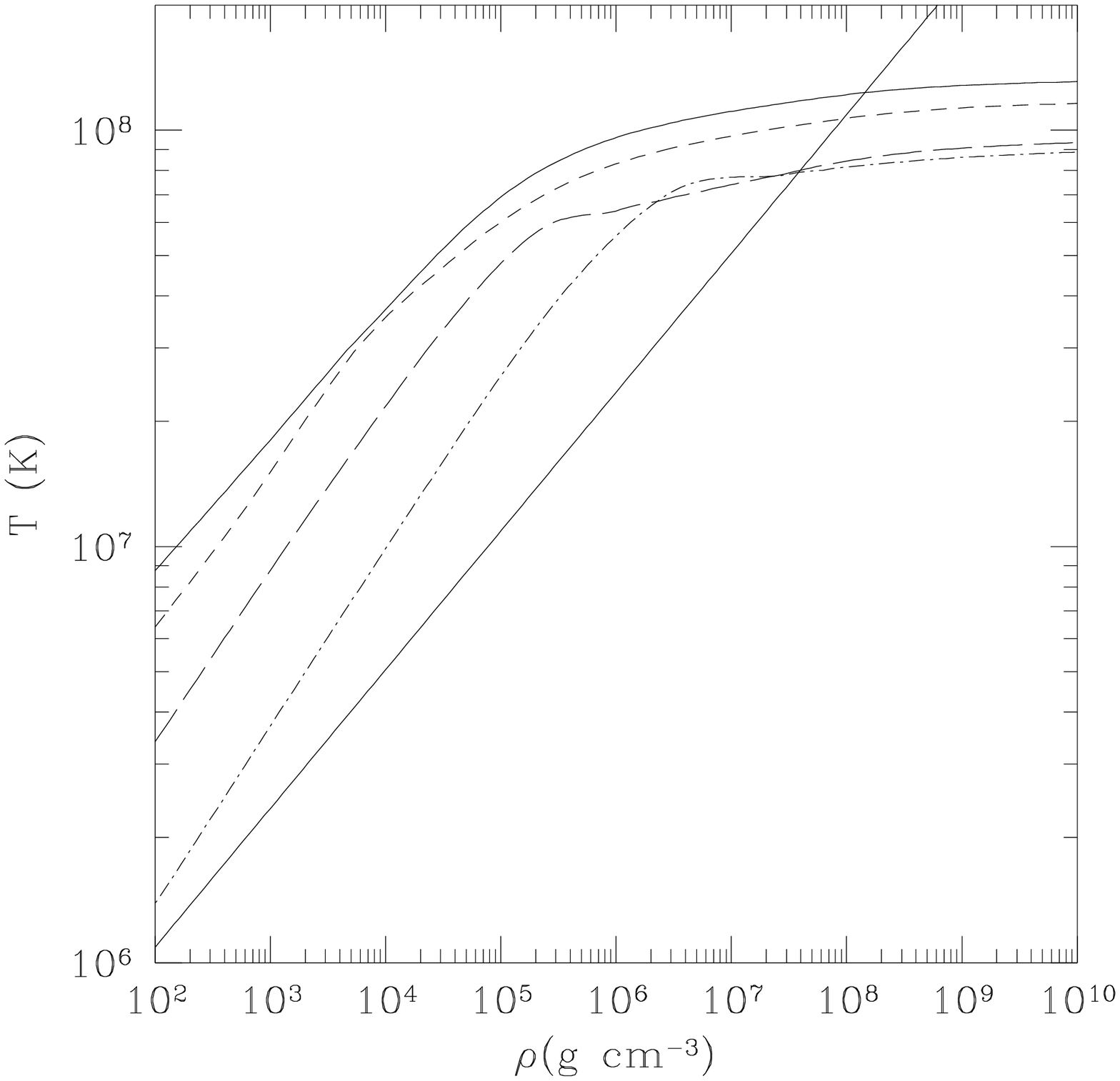}{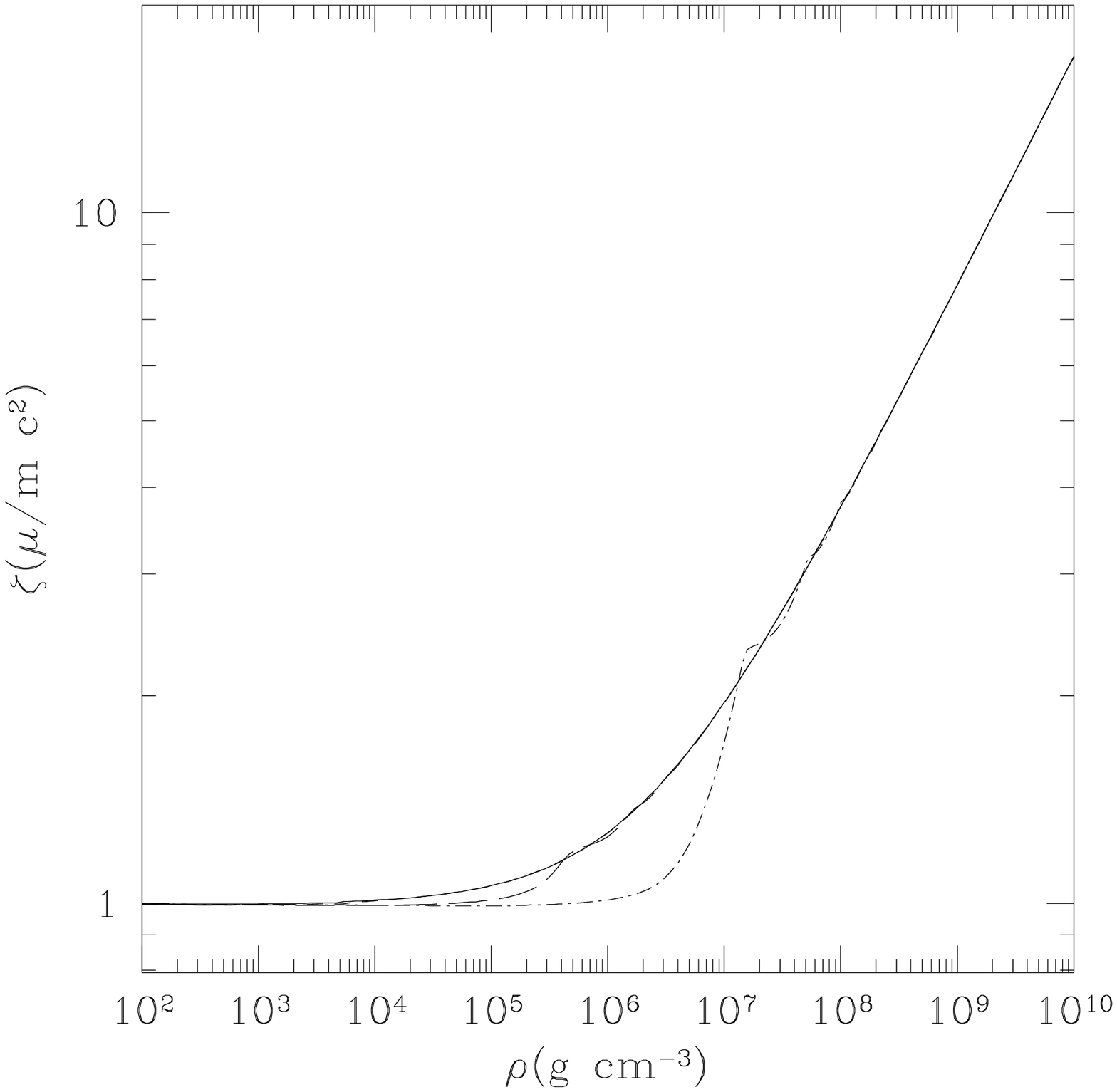}

\plottwo{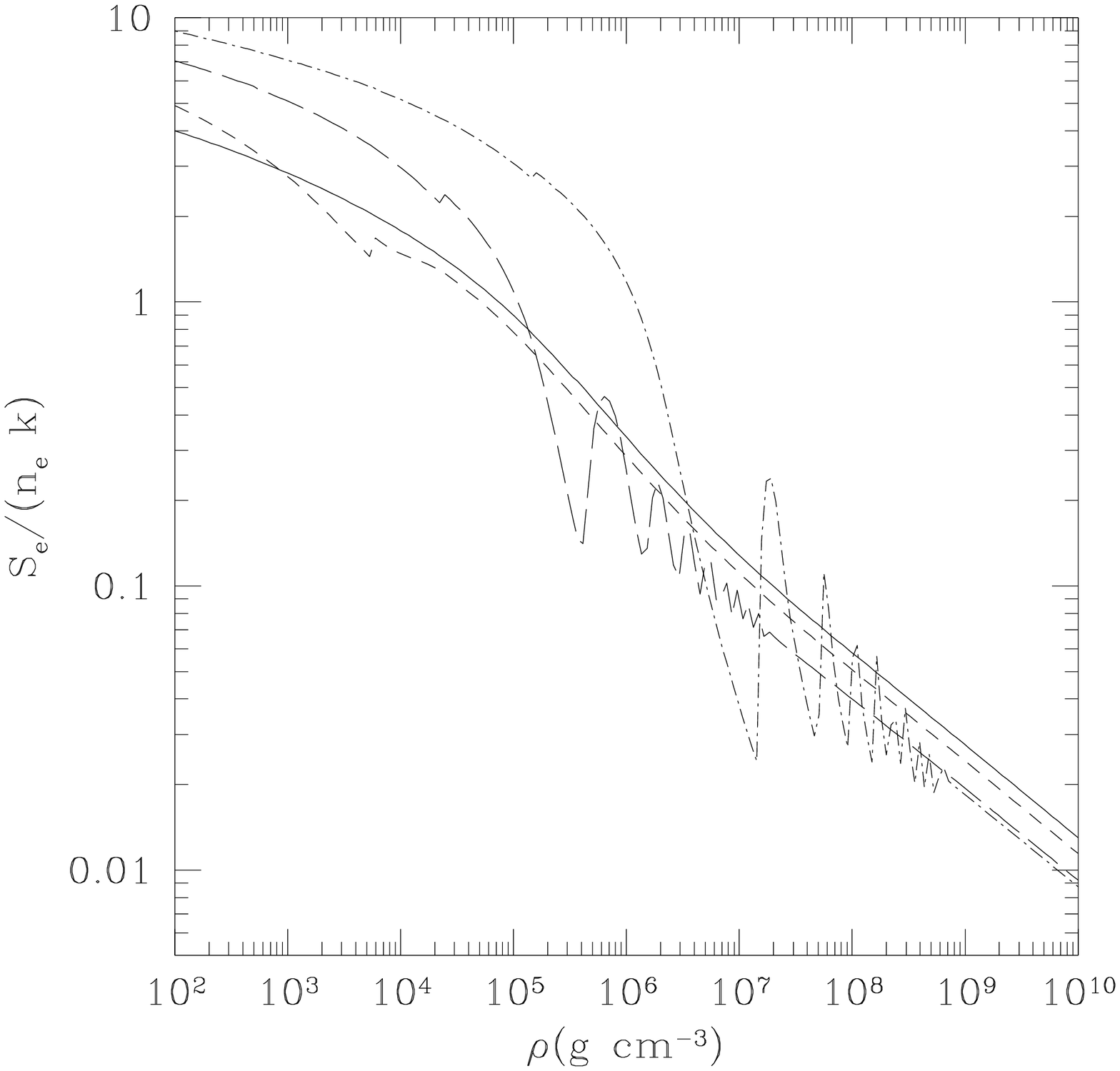}{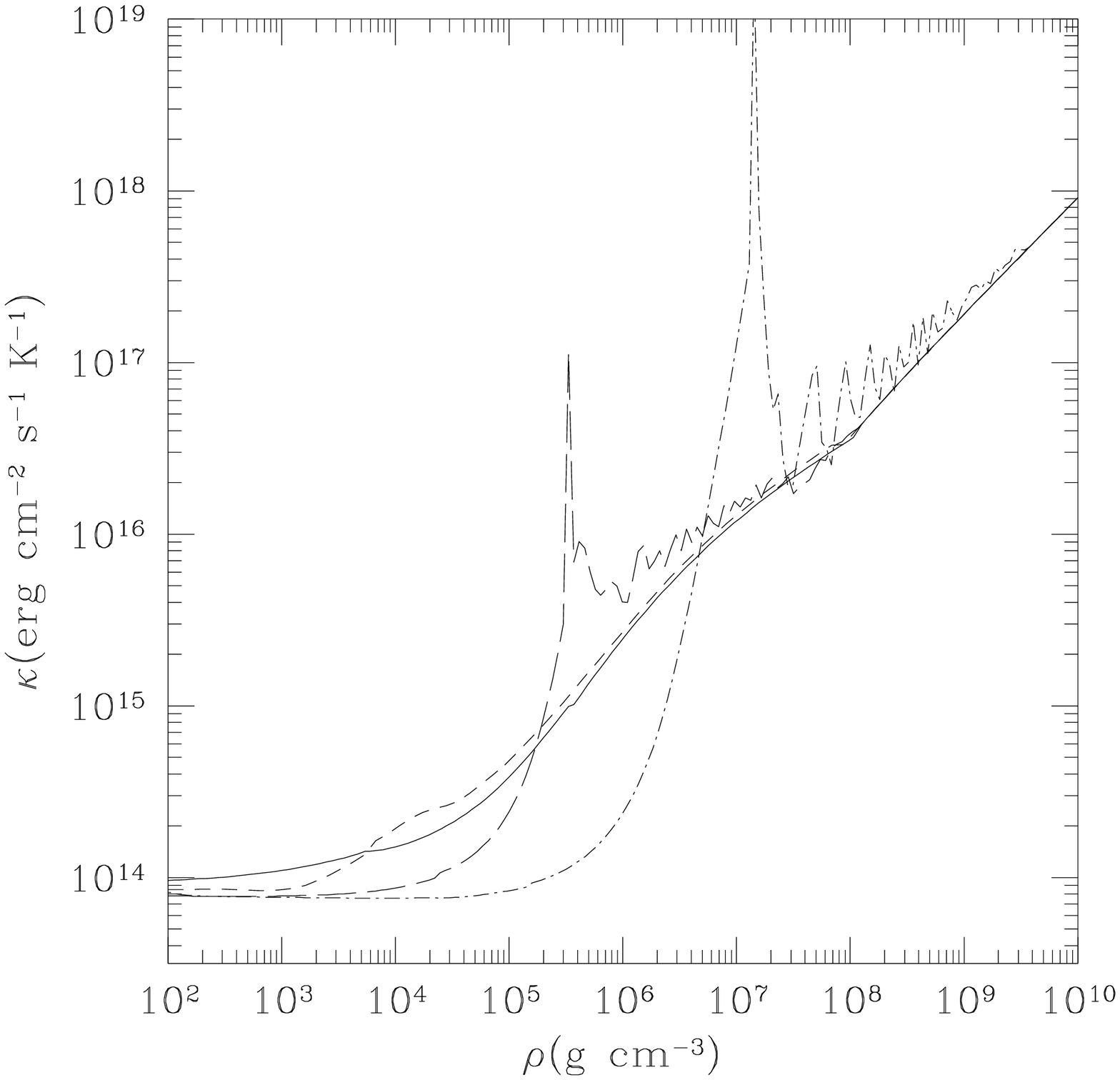}
\caption[The thermal structure for a radial field at an effective temperature of $10^6$~K]{
The thermal structure for a radial field at an effective temperature of $10^6$~K.  The solid curve
 traces the solutions for $B=0$, the short dashed curve gives $B=10^{12}$
G, the long dashed curve is $10^{13}$~G and the dot-dashed curve follows
the $B=10^{14}$~G solution.  The solid line traces the solid-liquid
phase transition in the $\rho-T$ plane.}
\label{fig:ts60para}
\end{figure}}

\figref{ts60para} depicts the temperature, chemical potential, entropy
and thermal conductivity as a function of density through the crust at
$T_\rmscr{eff}=10^6$~K.  The small discontinuity in the value of $S_e$
at low densities occurs when the integrator switches from using the
nondegenerate, nonrelativistic expression for $\zeta(n,\tau;\beta)$ to
numerically solving for $\zeta$.  This discontinuity does not affect the
integration through the nondegenerate regime. 

The run of electron entropy as a function of density or depth through
the envelope is not monotonic in the presence of a strong magnetic
field.  When one studies the total entropy, the nuclear contribution
weakens this effect, but the total entropy still does attain a
maximum as the first Landau level is being filled.  This entropy
inversion indicates that magnetized neutron star envelopes may be
convectively unstable.  However, a strong magnetic field also stabilizes
a material against convection (\cite{Chan61}).  To determine whether
convection is indeed important requires further study

Our values for the core temperatures using the \jcite{Hern84a}
conductivities are generally slightly higher for strong fields ($\sim
3 \%$) than those obtained by \jcite{Hern85} because we have varied
$\Lambda_{ei}$ to ensure continuity between the magnetized and the
unmagnetized conductivities.  A more substantial difference is
apparent in the value of the thermal conductivity in the nondegenerate
regime.  Because the thermal conductivity is nearly a power law in
this region, we expect the conductivity to be almost constant along a
solution here.  According to \jcite{Hern84b} for a unmagnetized
atmosphere, the conductivity along a solution in the nondegenerate
regime is
\be
\kappa = \frac{\alpha+\delta}{\alpha} \frac{F}{g_s} \frac{Y_e
k}{m_u} 
\label{eq:kappasolnndnb}
\ee
where $\alpha=2$ and $\delta=6.5$ for free-free scattering.  However,
in the magnetized case we obtain, 
\be
\kappa = \frac{\alpha+\delta-2}{\alpha} \frac{F}{g_s} \frac{Y_e
k}{m_u} .
\label{eq:kappasolnnd2}
\ee
The thermal conductivity along a solution in the nondegenerate regime
should be about 45 \%\ larger for an unmagnetized envelope than for a
magnetized envelope.  This effect is apparent in
Figures~\ref{fig:ts60para} and~\ref{fig:ts5565para}, but not in
Figures~7 through~9 of \jcite{Hern85}.  We find that the thermal
conductivity in the nondegenerate region is given precisely by
\eqref{kappasolnndnb} for the unmagnetized envelopes and by
\eqref{kappasolnnd2} for the magnetized ones.  The results of
\jcite{Hern85} for both magnetized and unmagnetized envelopes follow
\eqref{kappasolnndnb}.  We are not certain of the origin of this
discrepancy, but we suspect that it is due to inaccuracies in evaluating
the conductivities in the nondegenerate limit.  We expect that the
results for cool envelopes which depend sensitively on conductivity in
the nondegenerate 
region will differ between the work presented here and that of
\jcite{Hern85}.  Otherwise, the results for $T_\rmscr{eff}=10^6$~K
agree well with those of \jcite{Hern85}.

\figcom{\begin{figure}
\plottwo{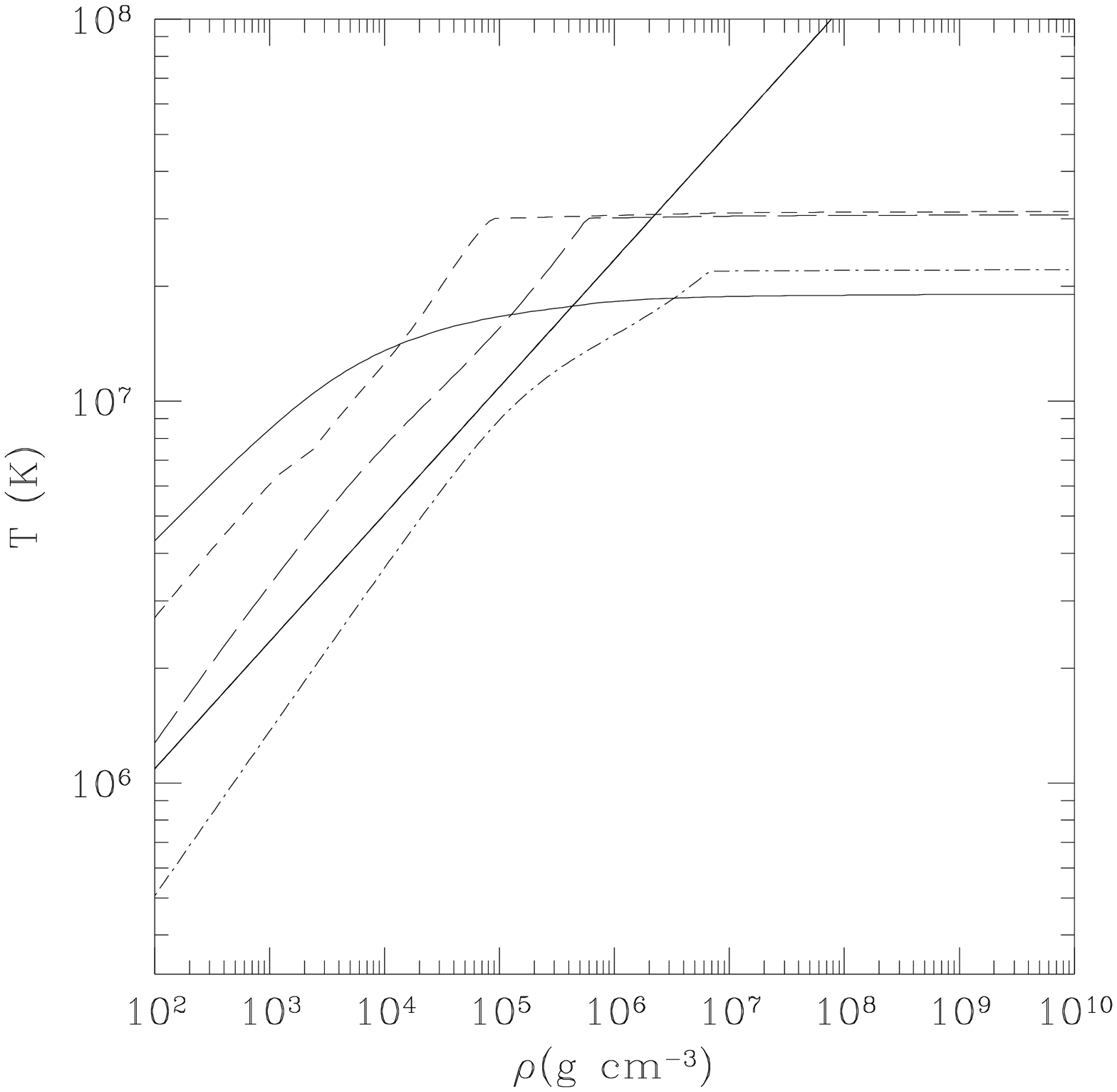}{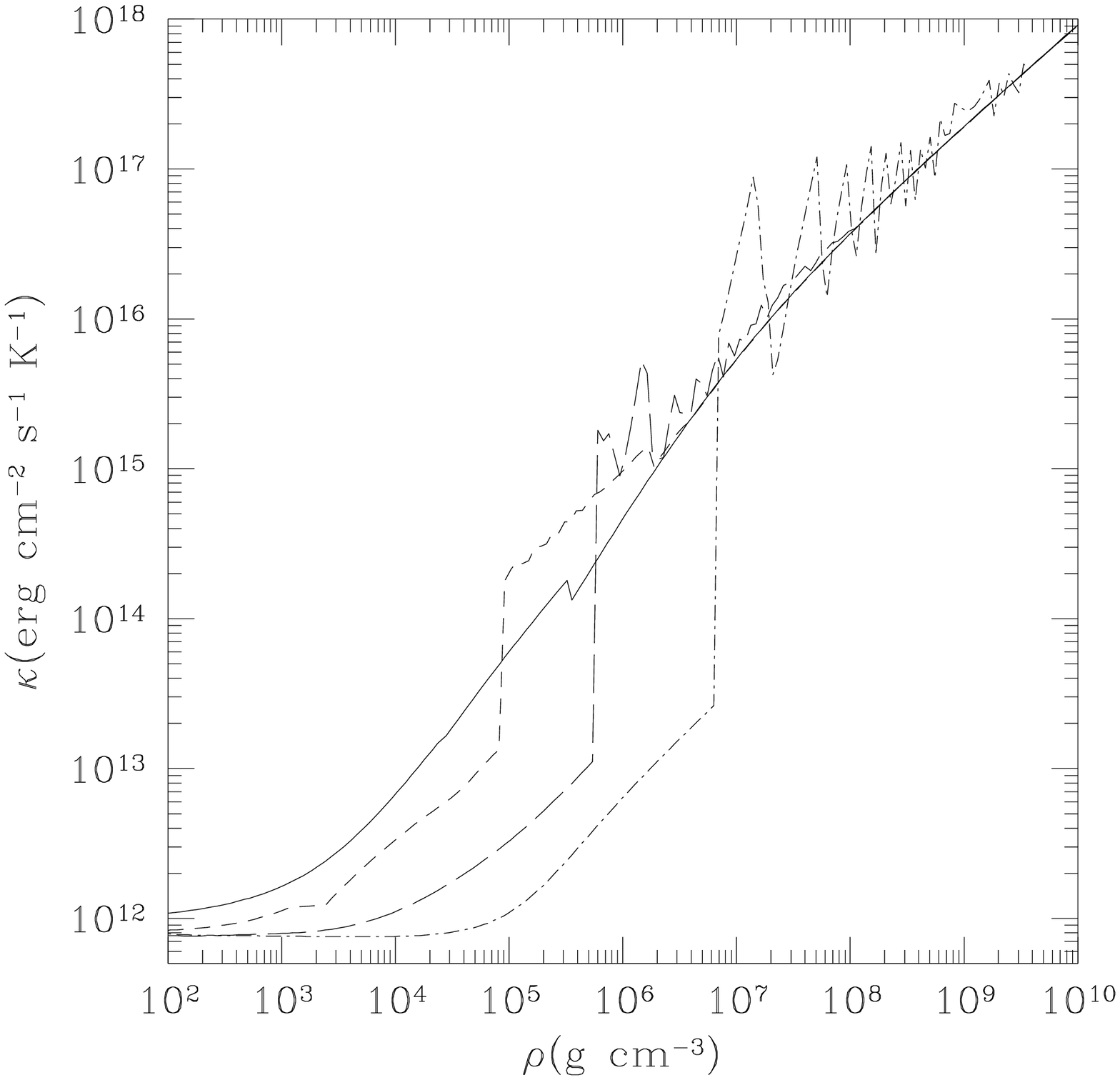}

\plottwo{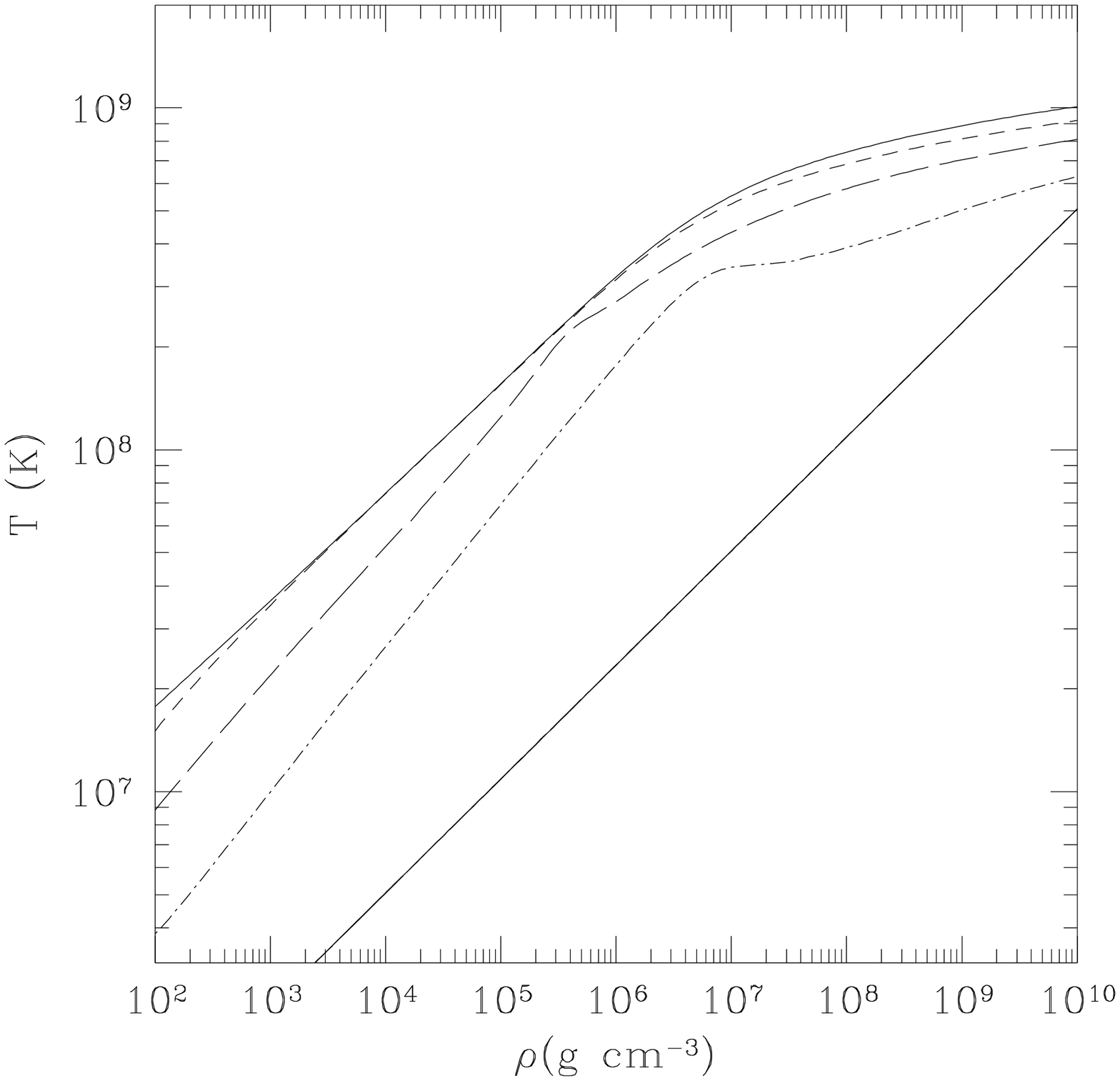}{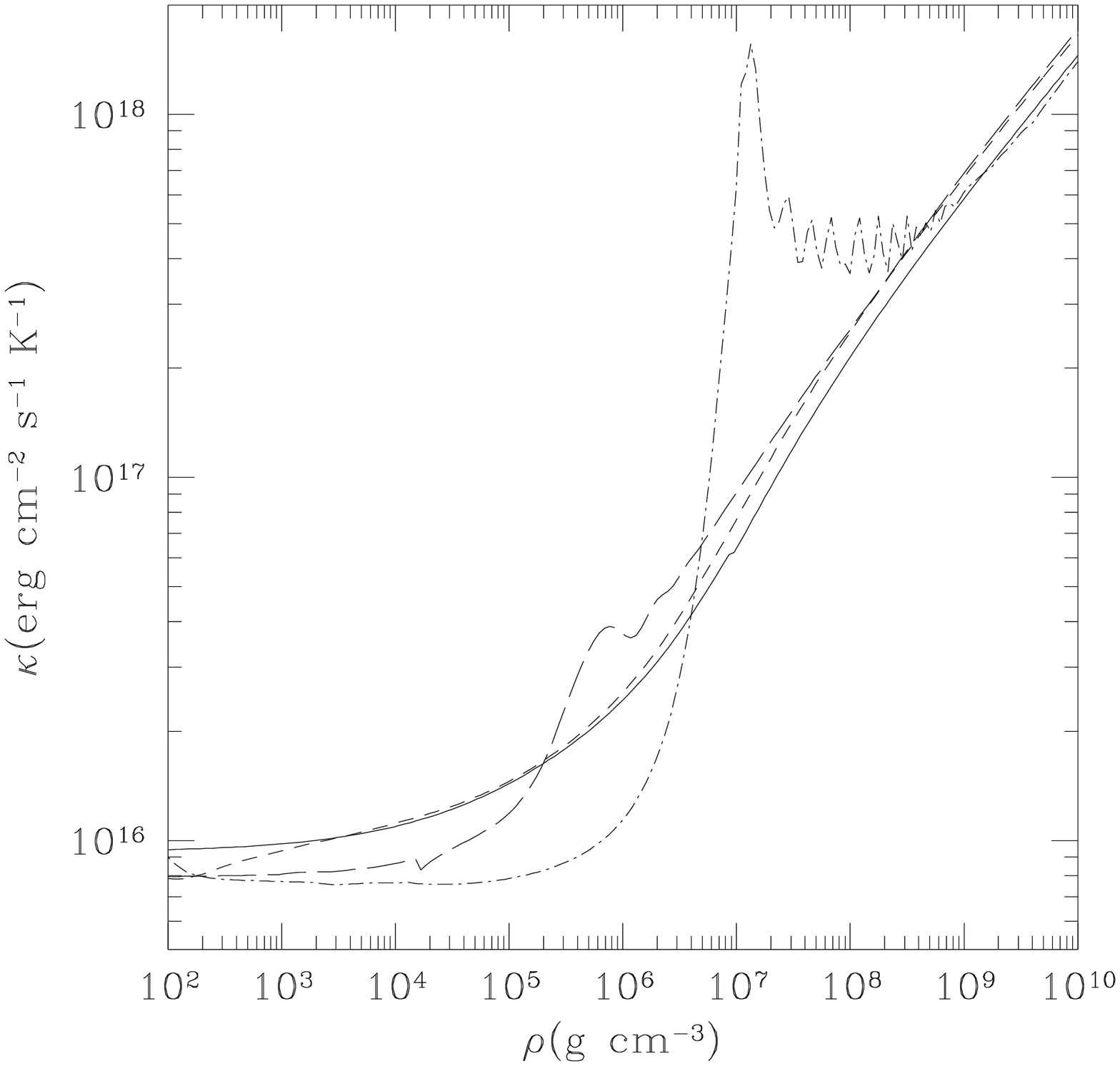}
\caption[The thermal structure for a radial field at effective
temperatures of $10^{5.5}, 10^{6.5}$~K]{
The thermal structure for a radial field at an effective temperature of
$10^{5.5}$~K (upper panels) and $10^{6.5}$~K (lower panels).  
The lines follow the solutions for the same field strengths as
 in \figref{ts60para}.}
\label{fig:ts5565para}
\end{figure}}

\figref{ts5565para} depicts the physical conditions of the envelope as a
function of density for effective temperatures of $10^{5.5}$~K (upper
panels) and $10^{6.5}$~K (lower panels).  In the colder envelope, the
oscillations of the thermodynamic quantities owing to the quantization
of the electron phase space is apparent.  At higher and lower effective
temperatures, the differences compared to the results of \jcite{Hern85}
are substantially larger.  We obtain the solutions with higher core
temperatures at $T_\rmscr{eff}=10^{5.5}$~K than \jcite{Hern85} did
because of the differences in the thermal conductivities in the
nondegenerate region.  Additionally, the relationship between the core
temperature and magnetic field strength is complicated at such low
effective temperatures.  Specifically we find that for $T_\rmscr{eff}
\leq 10^{5.6}$~K the core temperature for an unmagnetized envelope is
lower than in the magnetized case. 

At $T_\rmscr{eff}=10^{6.5}$~K the situation is reversed.  The envelopes
studied here tend to yield cooler core temperatures than those studied
by \jcite{Hern85}.  For the hot envelopes with $B \lesssim 10^{13}$~G,
the core temperature depends sensitively on the thermal conductivity in
the liquefied region where more than thirty Landau levels are filled.
In this region, unlike in \jcite{Hern85}, we have adjusted the value of
$\Lambda_{ei}$ to ensure continuity between the magnetized results and
the unmagnetized limit.  This yields slightly higher parallel
conductivities for the liquid state, and consequently lower core
temperatures.

In \figref{rt14anal}, we compare the $10^{14}$~G models with the
analytic models discussed in Paper~I at $T_\rmscr{eff} =
10^{5.5}, 10^6$ and $10^{6.5}$~K.  At $T_\rmscr{eff} = 10^6$~K, the
numerical model has a core temperature 11 \%\ cooler than the analytic
treatment.  The two approximations in the analytic model contribute
errors that partly cancel.  Since the analytic treatment assumes that
either photon or electron conduction operates, it underestimates the
conductivity in the semidegenerate region; consequently, in the
degenerate regime, the analytic envelope is slightly hotter, and the
resulting conductivity is larger than that of the numerical envelope at
the same density.  Above the density at which the first Landau level
fills, the analytic treatment effectively assumes that the conductivity
is infinite.  Because the core temperature depends most sensitively on
the conductivities in the semidegenerate region (\eg \cite{Gudm82}), the
net effect is that the numerical envelope at $10^6$~K yields slightly
higher core temperatures than the analytic treatment. 
\figcom{\begin{figure}
\plottwo{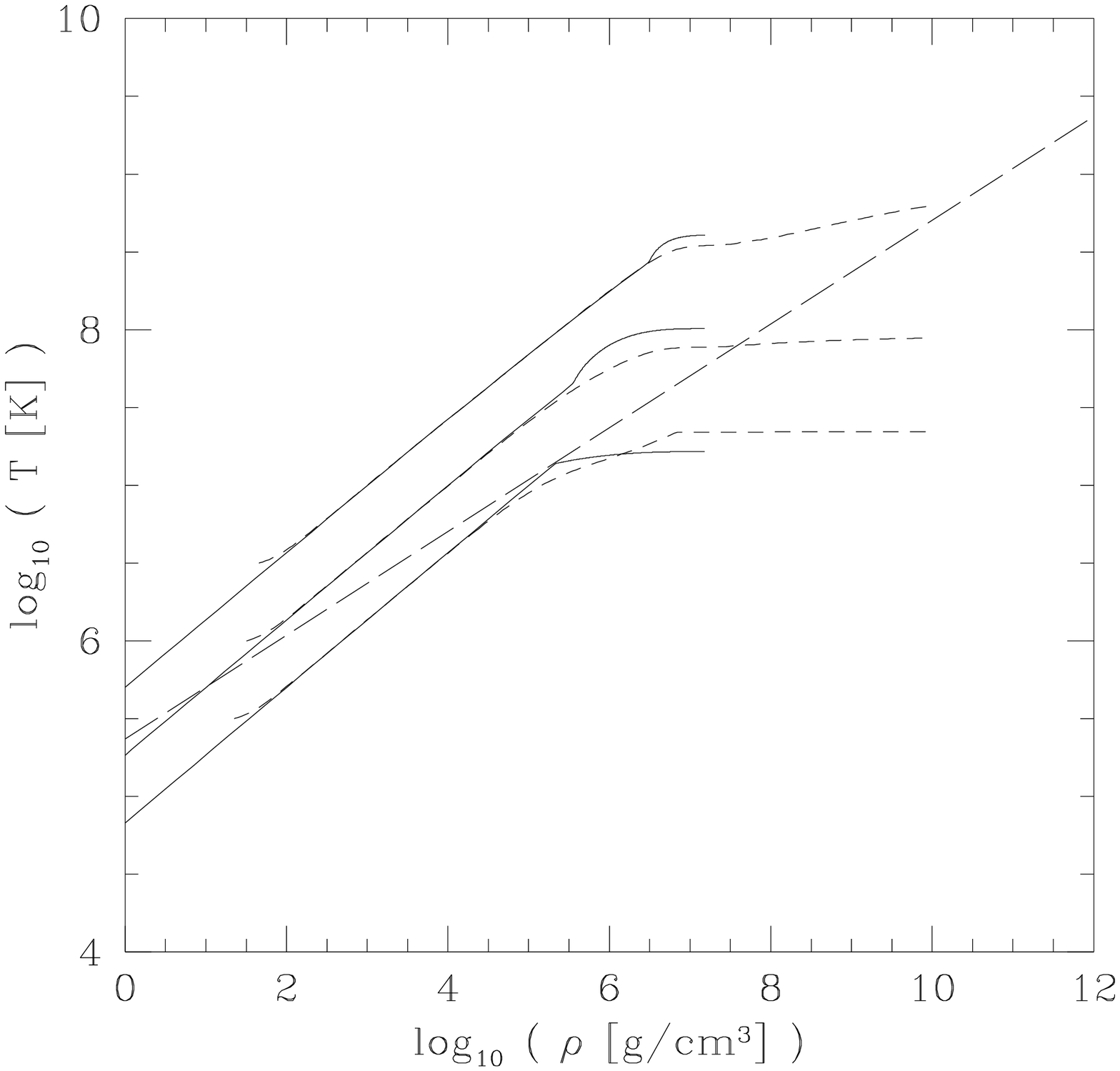}{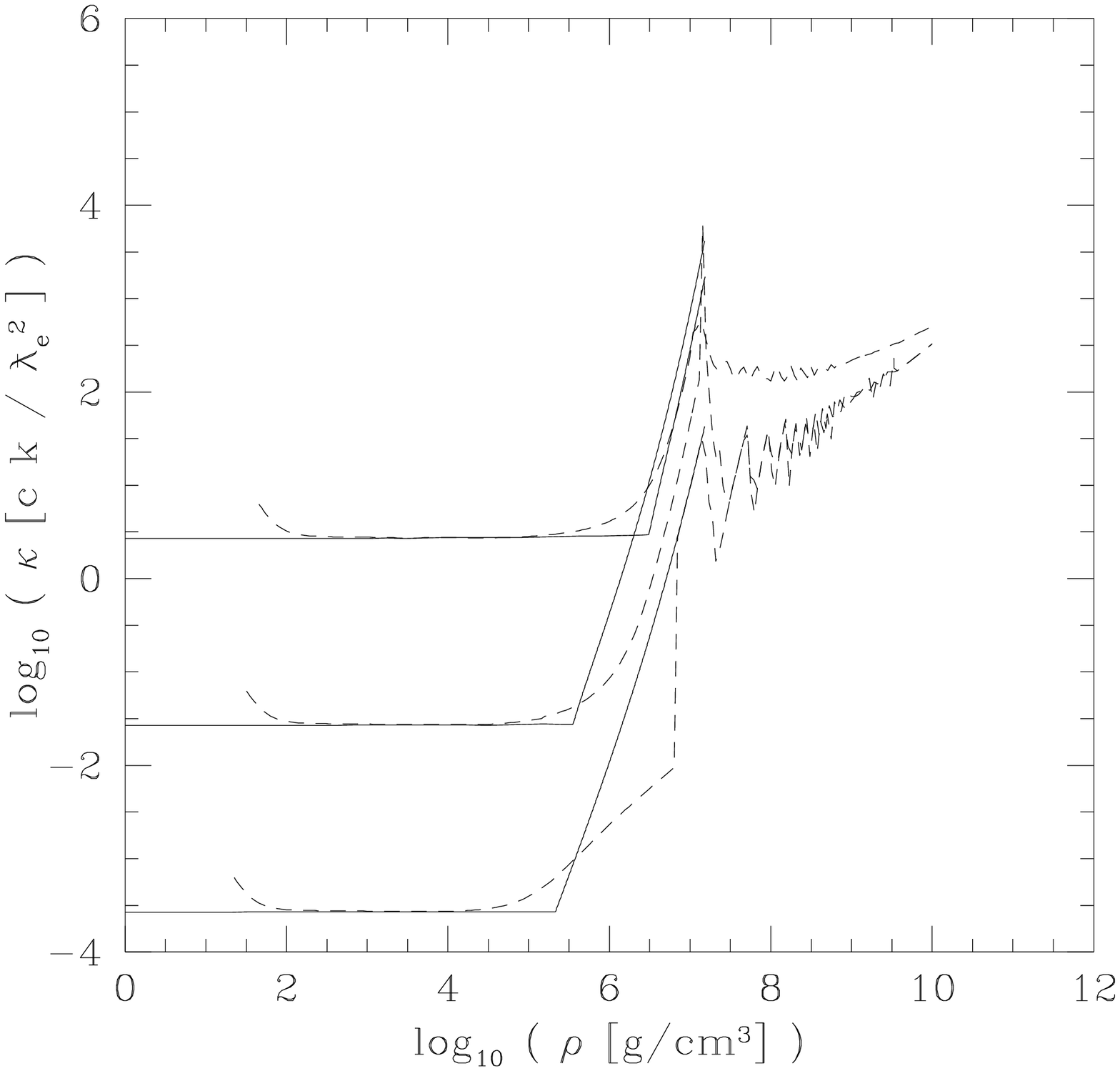}
\caption[Comparison of analytic and numerical envelope solutions.]{
Comparison of analytic and numerical envelope solutions for a radial
field. From top to bottom, the curves follow the solutions for $T_\rmscr{eff} =
10^{5.5}$~K, $10^6$~K and $10^{6.5}$~K.  The left panel depicts the
dependence of temperature on density through the envelope of the
neutron star.  The right panel gives the run of conductivity with
density.  The solid lines trace the analytic solutions and the dashed
follows the numerical results.}
\label{fig:rt14anal}
\end{figure}}

At very low densities, the conductivity for the numerical solutions is
higher than the constant conductivity of the analytic solution, because
in the numerical treatment we start with the photospheric boundary
condition rather than the radiative zero solution.  The photospheric
boundary condition assumes that $\tau=2/3$ where $T=T_\rmscr{eff}$.
The zero solution puts $T=T_\rmscr{eff}$ at $\tau=4/3$; therefore, at a
given low density, the numerical solutions are hotter than the
analytic ones.  However, the numerical solutions quickly relax to the
radiative zero solution (\eg \cite{Schw58}).

For hotter and cooler effective temperatures, the cancellation among
the errors introduced by the approximations is far less exact.  At
$T_\rmscr{eff}=10^{5.5}$~K the numerical treatment yields a core
temperature 30 \%\ hotter than the analytic model.  At
$T_\rmscr{eff}=10^{6.5}$~K the contribution to the insulation of the
core from material with more than one Landau level filled is
substantial and the core temperature estimate is 50 \%\ higher for the
numerical models than using the analytic treatment.

\paragraph{Perpendicular Conduction.}

Even classically, conduction perpendicular to the magnetic field is
affected dramatically by a strong magnetic field.  We again examine 
several magnetic field strengths.  \figref{ts55perp} depicts the
results for an effective temperature of $10^{5.5}$~K for perpendicular 
conduction.  The core temperature for a fixed effective temperature
varies dramatically with the magnetic field strength.  The bulk of the
effect is classical in nature, the magnetic field deflects the
electrons from carrying heat away from the surface.
\figcom{\begin{figure}
\plottwo{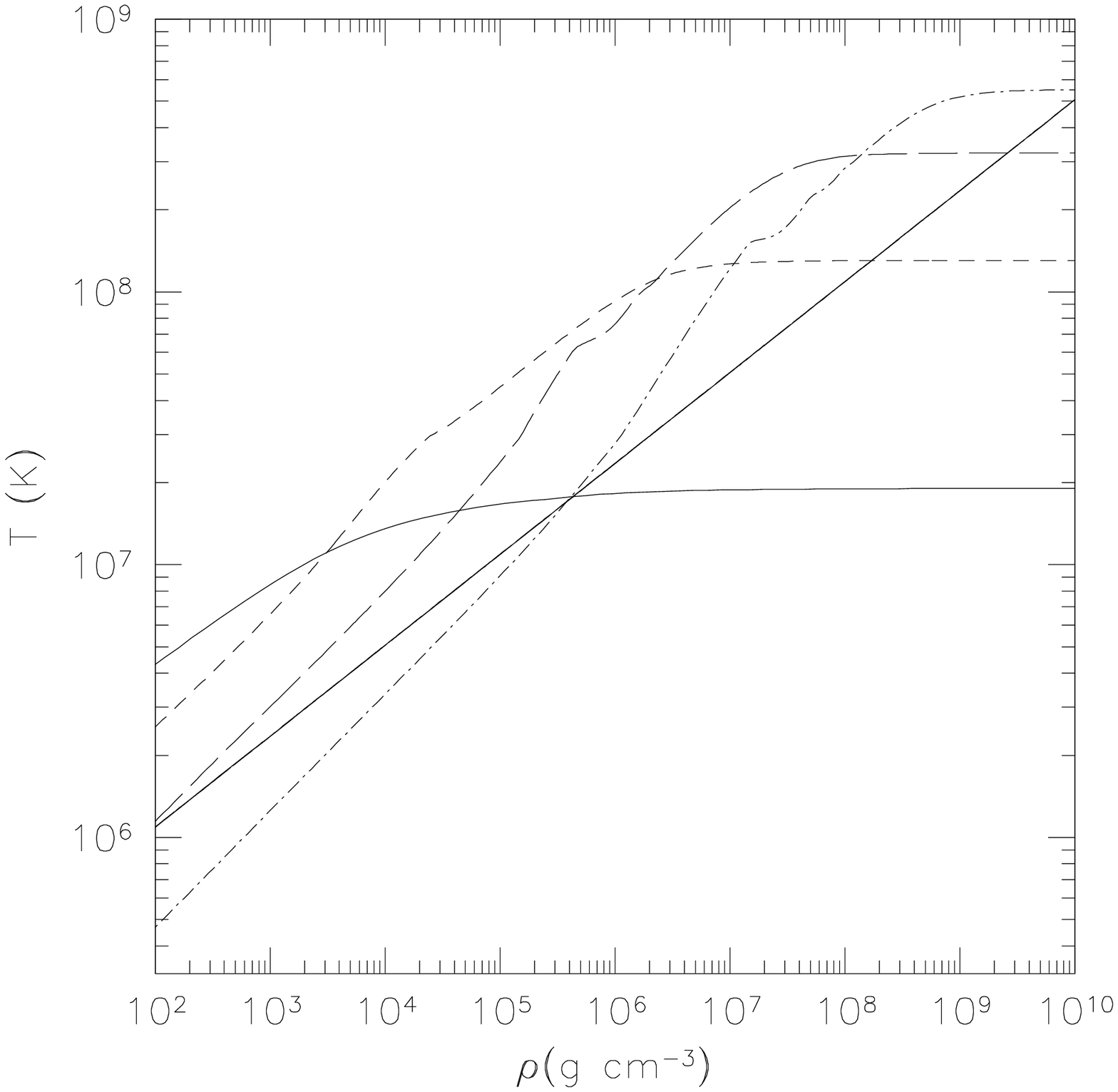}{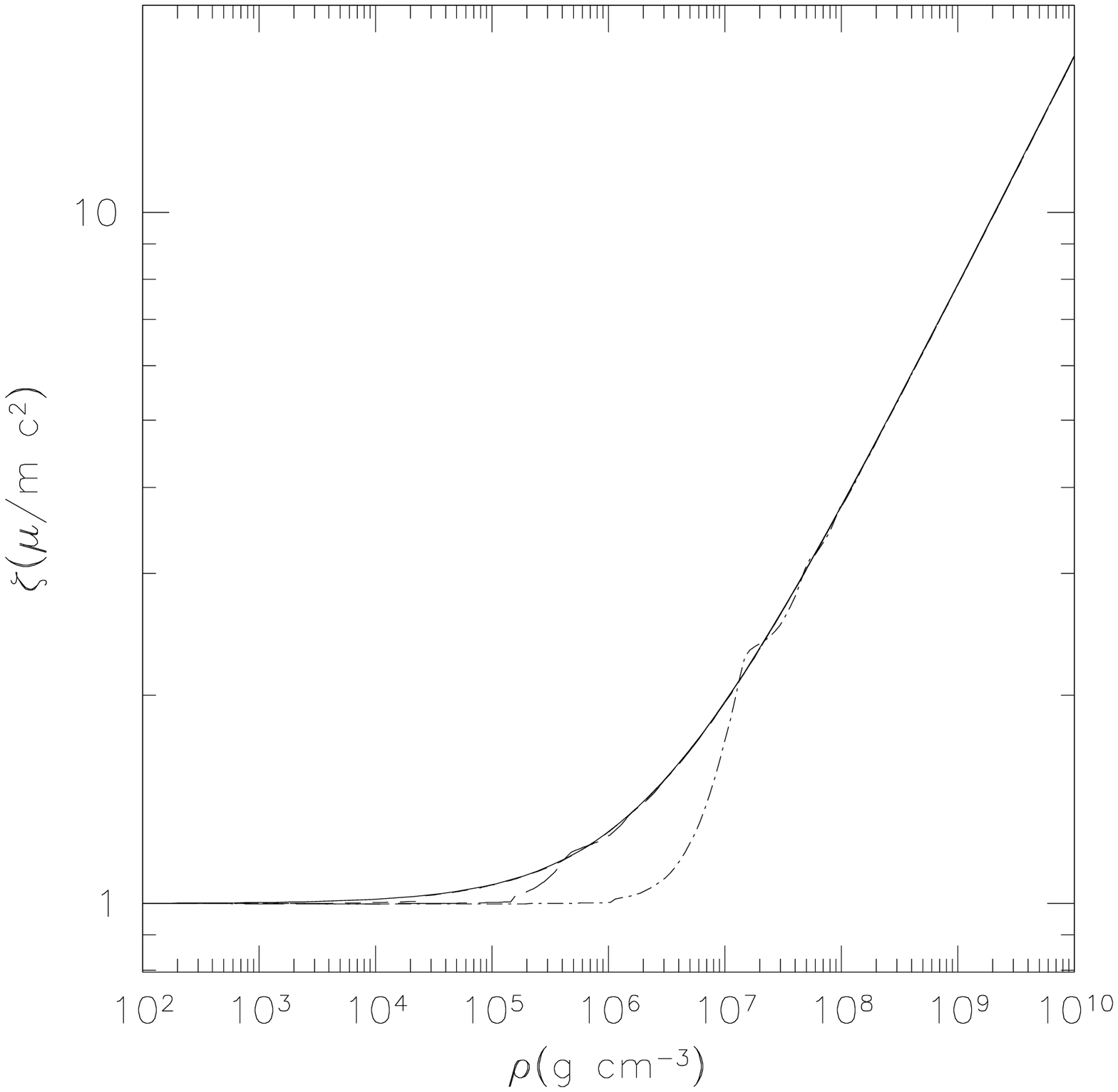}

\plottwo{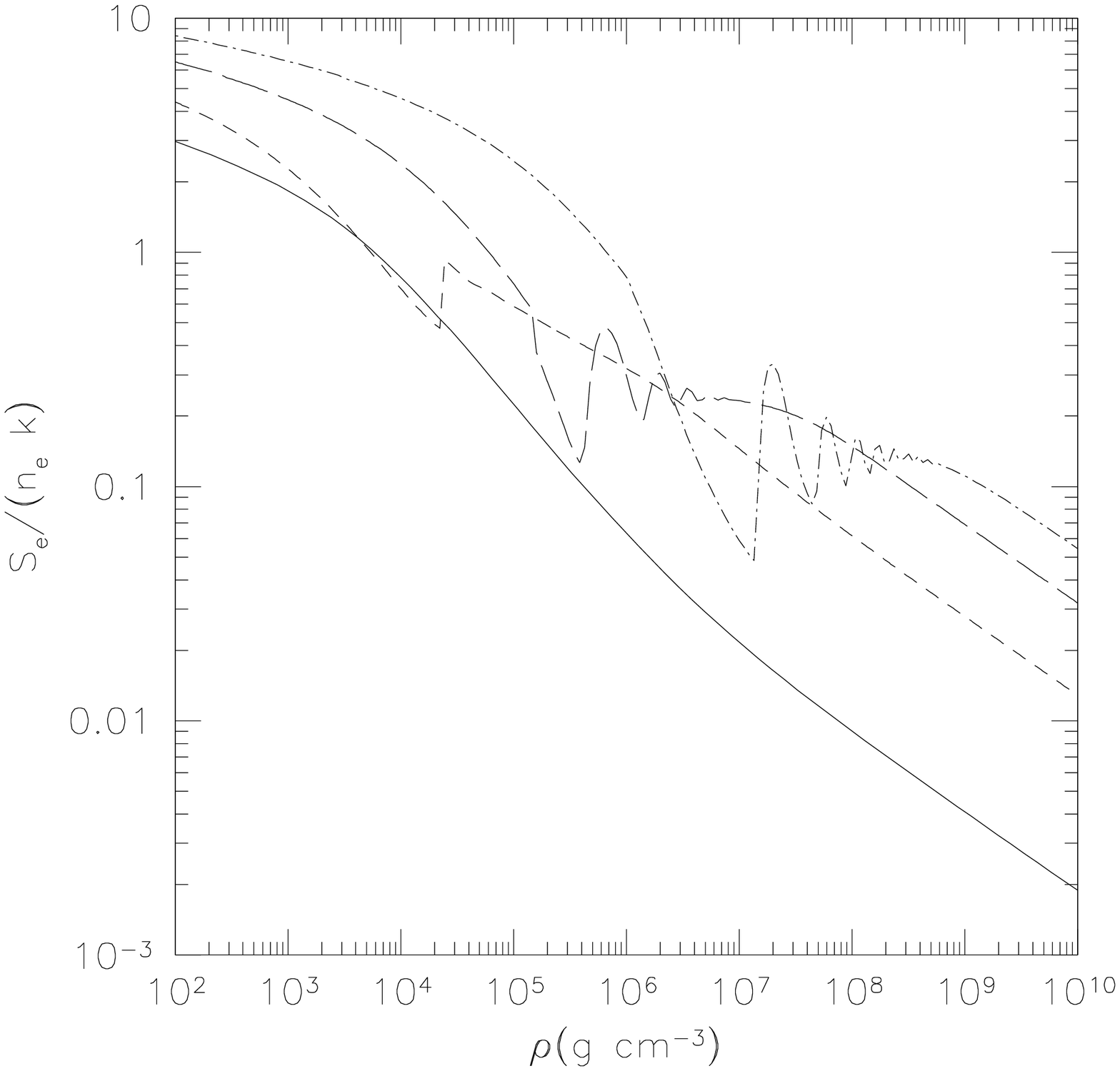}{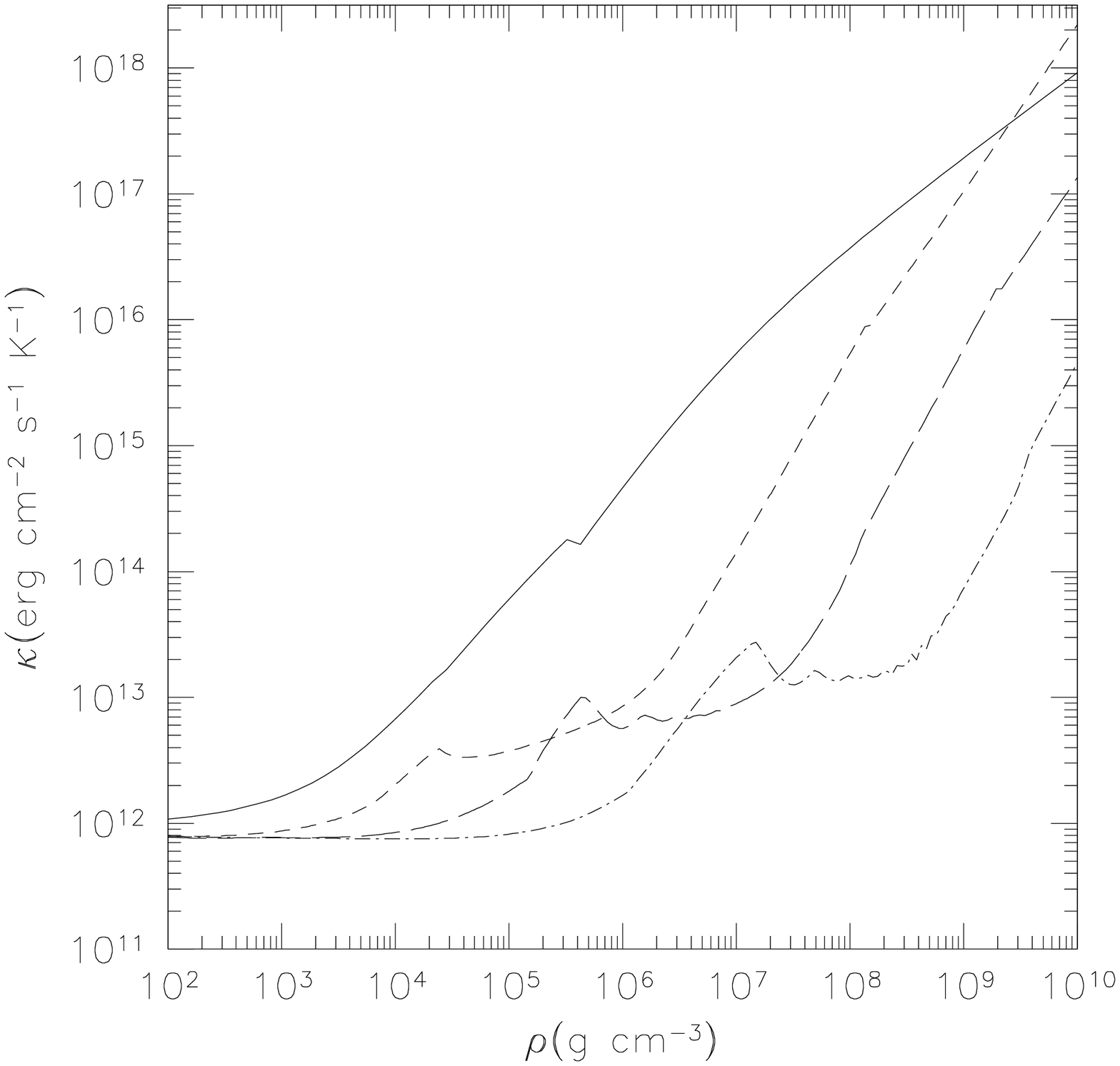}
\caption[The thermal structure for a tangential field at an effective
temperature of $10^{5.5}$~K]{ The thermal structure for a tangential
field at an effective temperature of $10^{5.5}$~K.  The lines follow the
solutions for the same field strengths as in \figref{ts60para}.}
\label{fig:ts55perp}
\end{figure}}

The quantization of the electron phase space is also manifest for the
perpendicular case.  Because the function $Q(\nu,\beta)$ diverges as a Landau
level begins to fill, the conductivity increases dramatically near the
start of each Landau level, and the run of temperature exhibits plateaux
at these densities.  As we shall find in the next subsections,
perpendicular transport cannot be neglected an important range of
effective temperatures and magnetic field strengths. 

\subsection{Angular Dependence}

To examine the angular dependence of the flux transmitted through a
uniformly magnetized envelope, we will take two routes.  First, we use
the method demonstrated in Figure~7 of Paper~I and vary the effective
temperature as a function of angle to shoot toward a fixed core
temperature.

\figcom{\begin{figure}
\plottwo{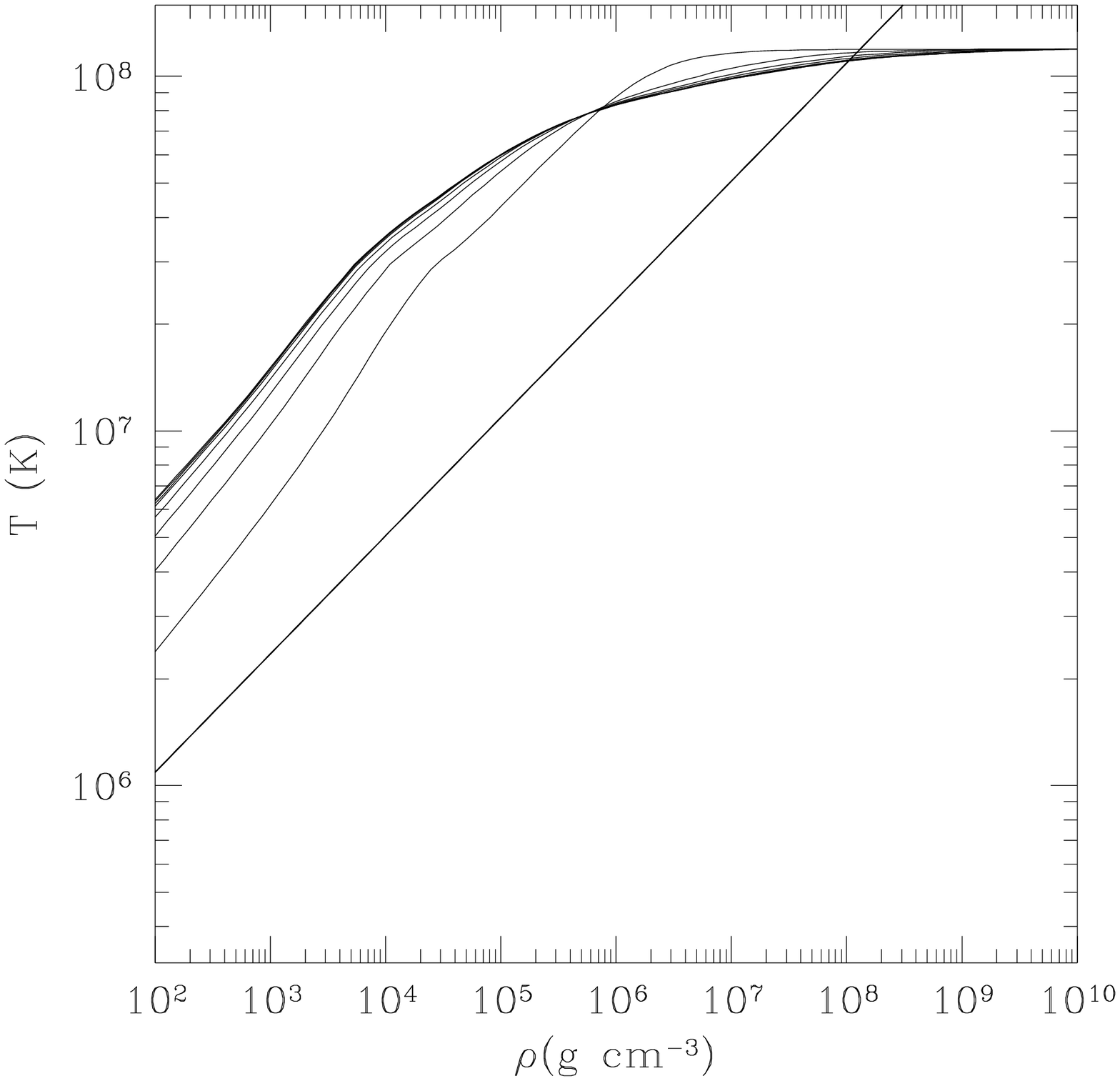}{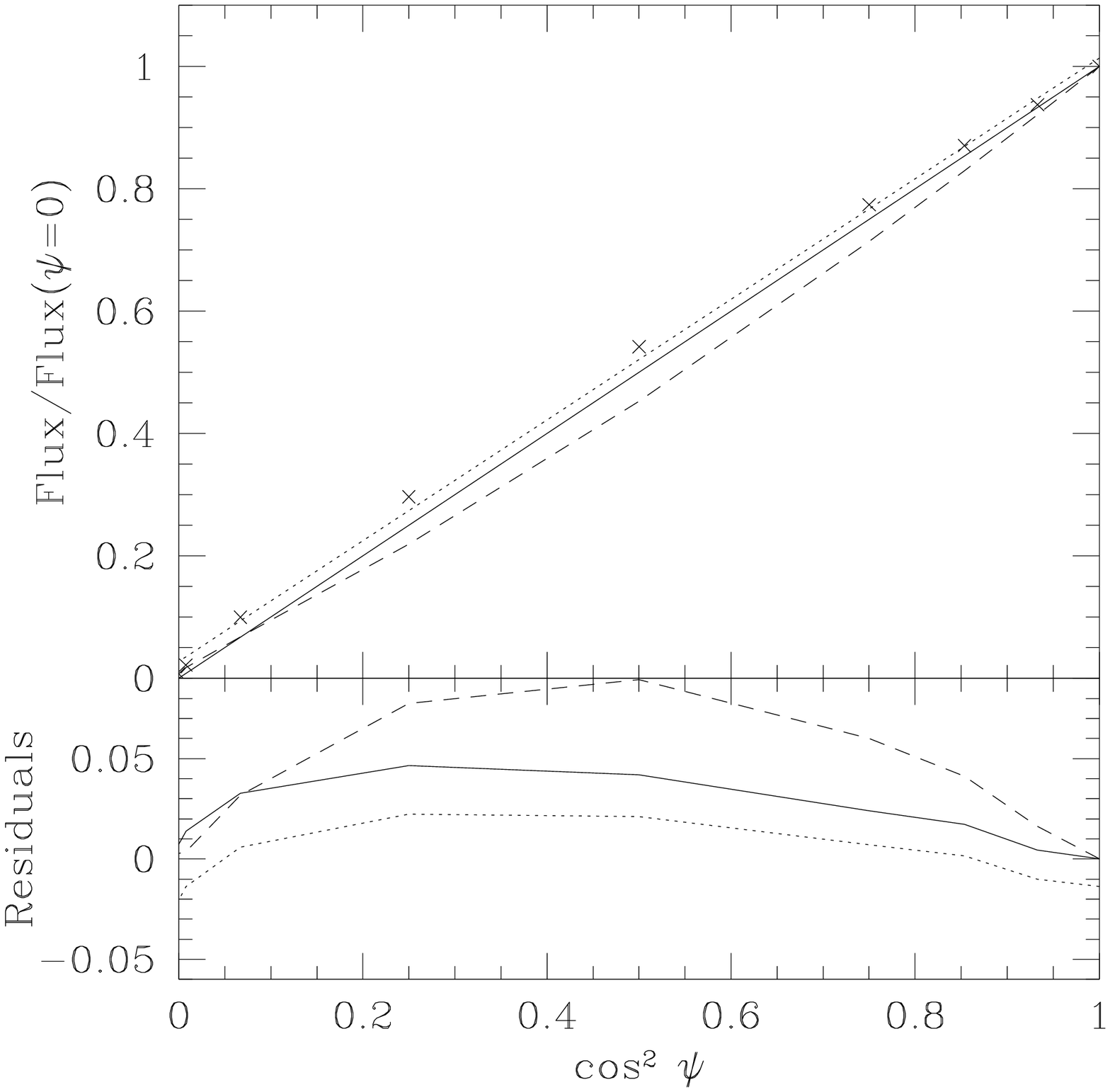}

\plottwo{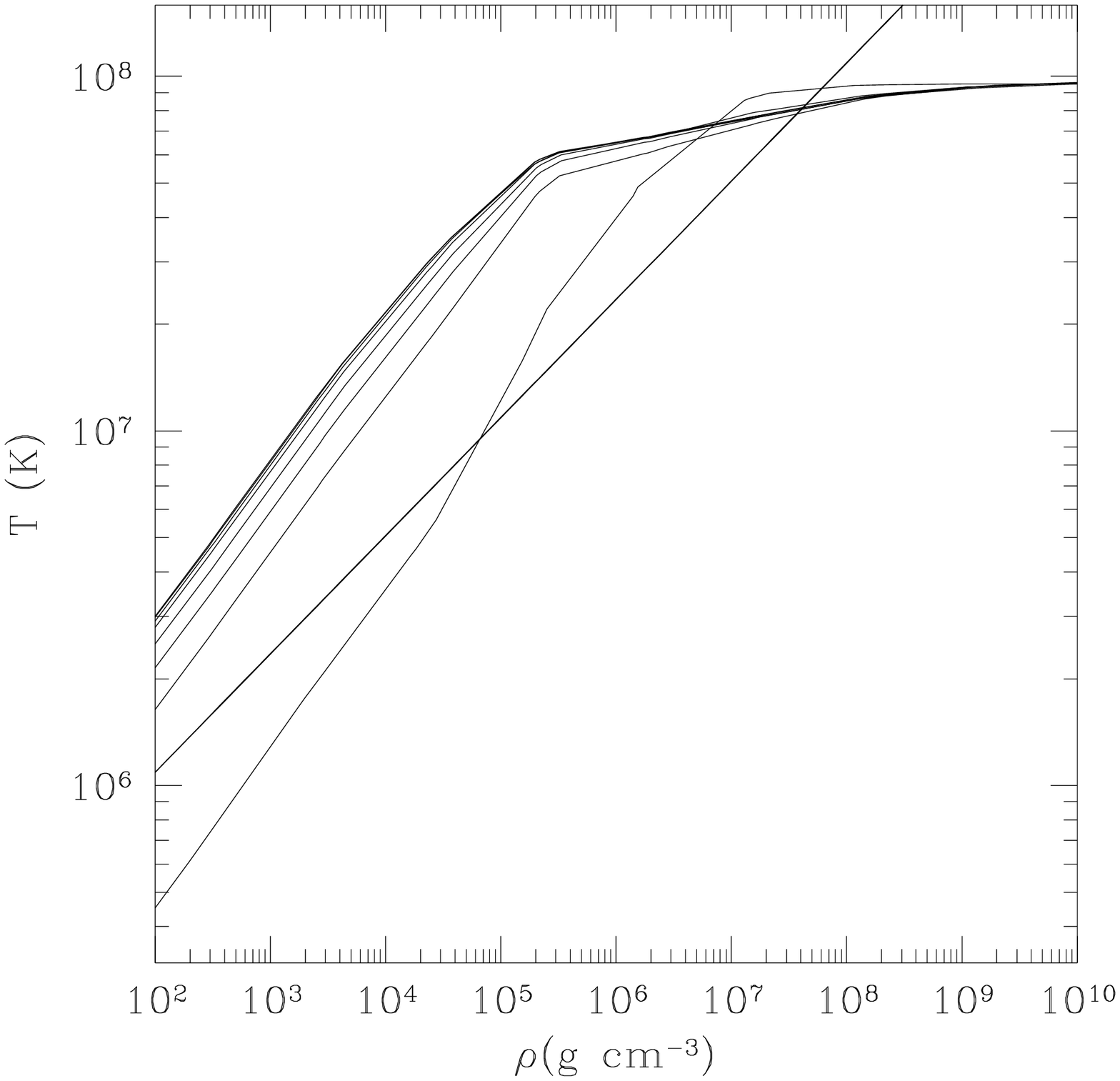}{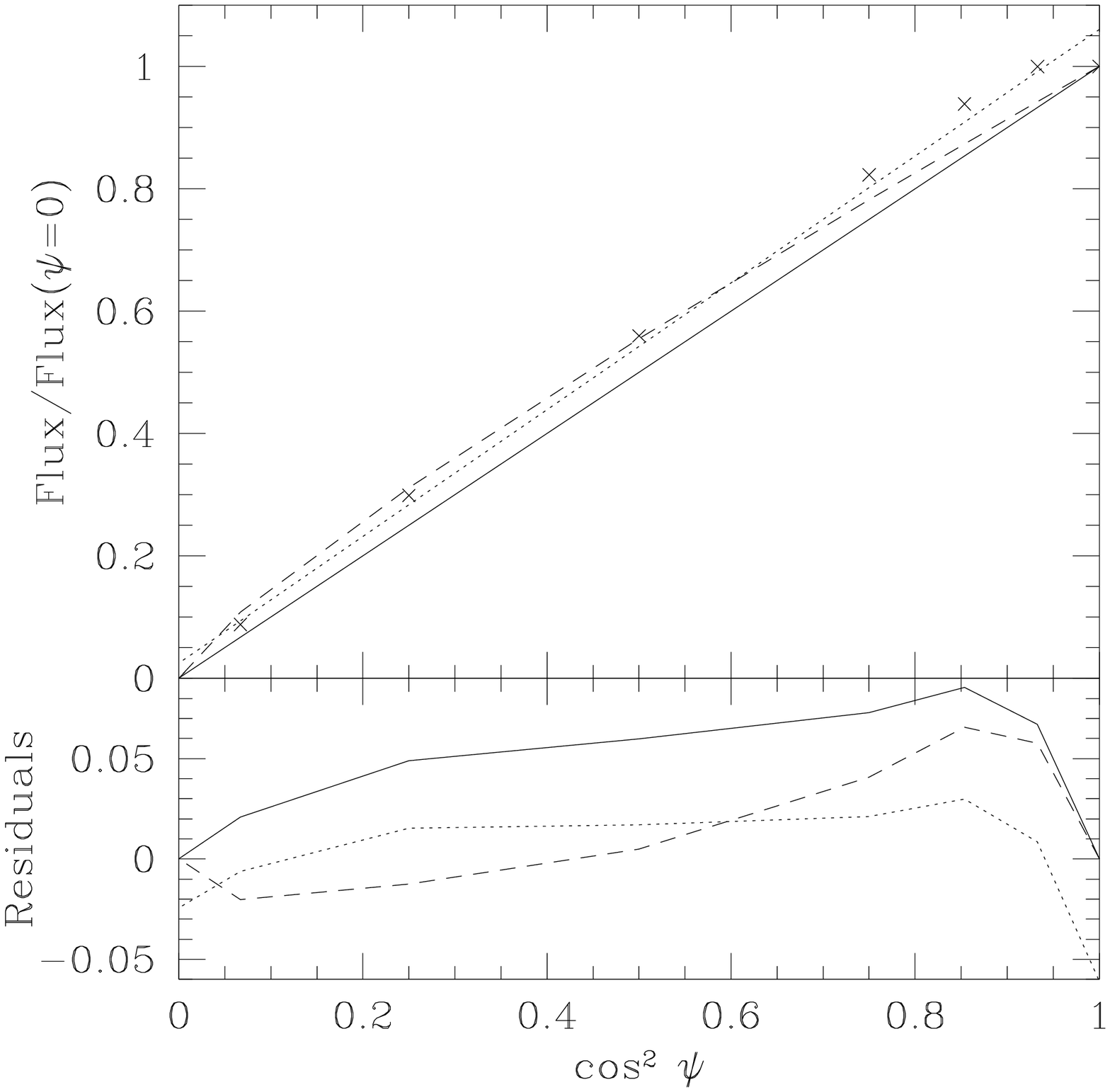}
\caption[Results of a numerical two-dimensional calculation for
$B=10^{12}, 10^{13}$~G]{
Results of a numerical two-dimensional calculation for $B=10^{12},
10^{13}$~G at $T_\rmscr{eff}=10^6$~K for $\psi=0$.  The upper panels
present the results for the weaker field strength.  The left panels
give $T(\rho)$ for the various models.  The right panels compare the
flux distribution (crosses) with the $\cos^2 \psi$ rule.  The lower
solid line gives the $\cos^2 \psi$ rule and the upper dotted line traces
the best fit model of the form $a \cos^2 \psi + b \sin^2 \psi$.  
Here, $a=1.02$ and $b=0.0264$ for $10^{12}$~G, $a=1.06$ and $b=0.0245$ for
$10^{13}$~G. The dashed line traces the results of
\jcite{Scha90b} (extrapolated using a power law to  $B=10^{13}$~G).  }
\label{fig:an1360}
\end{figure}}

\figref{an1360} gives the outcome of this procedure.  Both the $\cos^2
\psi$ rule and the best fit $a \cos^2 \psi + b \sin^2 \psi$ match the
results to within 10 \%\ of the total flux.  \jcite{Scha90b} presents
the results of a set of two-dimensional calculations with a fitting
function
\be
\frac{T_\rmscr{eff}(\psi)}{T_\rmscr{eff}(0)} = \chi(\psi) = 
\chi \left ( 90^\circ \right ) 
+ \left [ 1 - \chi \left ( 90^\circ \right ) \right ] \cos^\alpha \psi.
\label{eq:Schamod}
\ee
\jcite{Scha90b} studies field strengths up to $10^{12}$~G, so we assume
that the parameters $\alpha$ and $\chi \left ( 90^\circ \right )$ for
$T_\rmscr{eff}=10^6$~K follow a power law in the field strength for
stronger fields.  For $B=10^{13}$~G, we obtain $\alpha=0.48$ and $\chi
\left ( 90^\circ \right ) = 0.10$.  This model is traced by the dashed
line in \figref{an1360} and agrees to within 6.5 \%\ of our results.

Bolstered by the success of the $\cos^2\psi$ rule, we perform a second
test in which the flux along the surface varies as $\cos^2 \psi$ and
determine by how much the core temperature changes for several models.
\figref{tcanggen} depicts the results of these calculations.  A
horizontal line for a given set of calculations would indicate adherence
to the $\cos^2\psi$ rule.  Generally, the largest departure from the
$\cos^2\psi$ rule is where the heat is transmitted at large angles to
the magnetic field direction. 
\figcom{\begin{figure}
\plotonesc{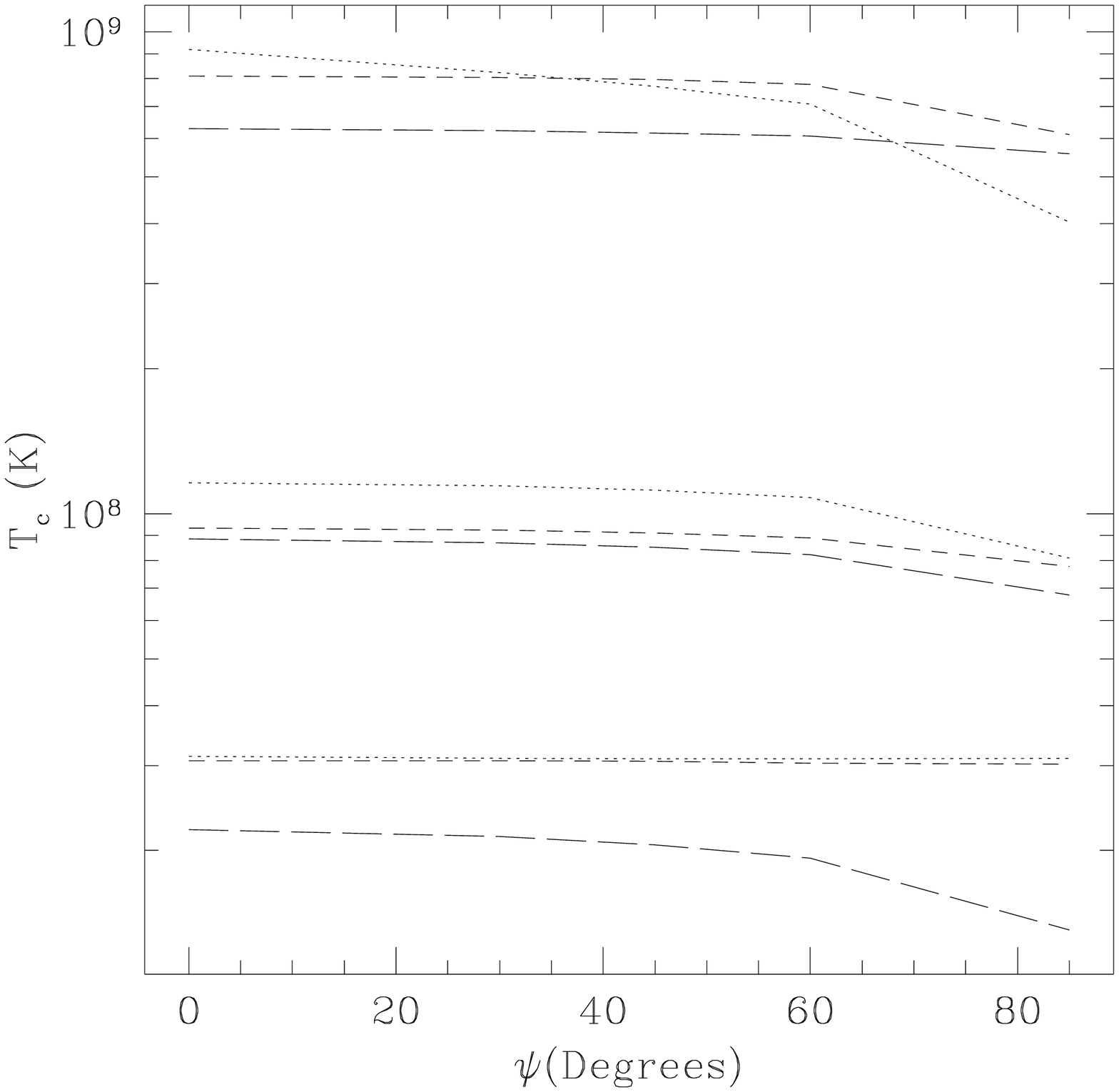}{0.7}
\caption[{The core temperature as a function of angle for fluxes that
follow the $\cos^2\psi$ rule}]{The core temperature as a function of
angle for fluxes that follow the $\cos^2\psi$ rule.  The lower lines are
for $T_\rmscr{eff}(\psi=0)=10^{5.5}$~K, and middle lines follow the
$10^6$~K solutions, and the upper lines trace the $10^{6.5}$~K results.
The lines follow the solutions for the same field strengths as in
\figref{ts60para}.  }
\label{fig:tcanggen}
\end{figure}}

We define a figure of merit for each value of the flux at $\psi=0$ and
magnetic field strength,
\be
\Upsilon_u = \frac{1}{N} \sum_{i=1}^N \left | \frac{T_c(\psi_i) - T_c(0)}{T_c(0)} \right |
\ee
where $u$ signifies an unweighted summation.

A large value of $\Upsilon_u$ indicates that conductivity
perpendicular to the field lines is significant in determining the
core temperature given the transmitted flux.  \tabref{upsilonBT}
depicts the results for several field strengths and effective
temperatures.  We find two trends.  For weak fields, the low effective
temperature solutions follow the $\cos^2\psi$ rule more closely than
hotter envelopes.  For strong fields, the trend is reversed.  
\figcom{\bt
\caption{Values of $\Upsilon$ in percent as a function of magnetic
field and effective temperature}
\label{tab:upsilonBT}
\begin{tabular}{c|rrr|rrr}
      & \multicolumn{3}{c|}{$\Upsilon_u$} 
      & \multicolumn{3}{c}{$\Upsilon_L$} \\
      & \multicolumn{3}{c|}{$\log T_\rmscr{eff}$} 
      & \multicolumn{3}{c}{$\log T_\rmscr{eff}$} \\
B (G) & 
\multicolumn{1}{c}{5.5} &
\multicolumn{1}{c}{6.0} & 
\multicolumn{1}{c|}{6.5} &
\multicolumn{1}{c}{5.5} &
\multicolumn{1}{c}{6.0} & 
\multicolumn{1}{c}{6.5} \\ \hline
$10^{12}$ & 0.86 & 8.4 & 21. & 1.1 & 3.7  & 16. \\
$10^{13}$ & 0.74 & 4.9 & 6.1 & 0.54 & 2.4 & 1.9 \\
$10^{14}$ & 12.  & 7.2 & 3.5 & 7.0  & 4.0 & 2.0 
\end{tabular}
\et}

>From an observational point of view, the error in the total predicted luminosity of 
the object is more important.  We weight the residuals by $\cos^2\psi \sin\psi$.
This neglects gravitational lensing and assumes the field distribution is uniform.  We
define
\be
\Upsilon_L = 
\sum_{i=1}^N \cos^2\psi \sin\psi \left | \frac{T_c(\psi_i) - T_c(0)}{T_c(0)} \right |
\Biggr / \sum_{i=1}^N \cos^2\psi \sin\psi
\ee
where $L$ signifies a luminosity-weighted summation.  The values of
$\Upsilon_L$ tend to be smaller than those of $\Upsilon_u$ because the
weighting function is peaked at $\cos^2\psi=2/3$, approximately
$35^\circ$, where the departure from the $\cos^2\psi$ rule is small.
We find that for effective temperatures near $10^6$~K the $\cos^2\psi$
dependence is followed.  However, one must be wary in applying this
rule for cool strongly magnetized envelopes where the quantization of
the electron phase space increases the perpendicular conductivity
dramatically or hot weakly magnetized ones where the classical
relaxation time is no longer long compared to the relativistic
cyclotron frequency.

\subsection{Flux-Core-Temperature Relation}

\figref{coretemp0} depicts the flux-core-temperature relation.
We see that for very cool magnetized envelopes, the
relationship departs from a power law.  However, for $T_\rmscr{eff}
\geq 10^{5.7}$~K, the relationship is well fitted by a power law with
root-mean-square residuals of less than 3\%.
\figcom{\begin{figure}
\plotonesc{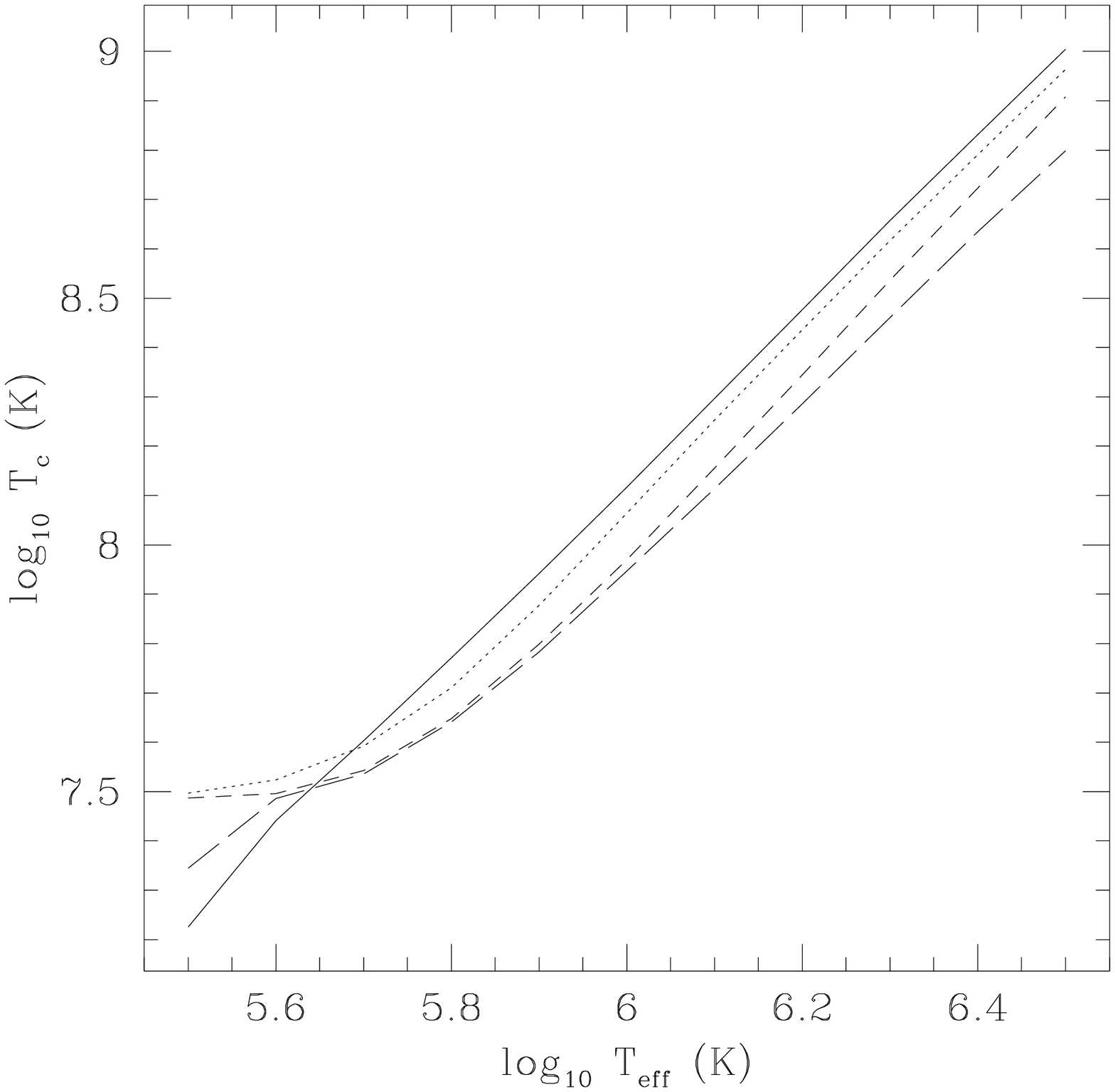}{0.7}
\caption[{The flux-core-temperature relation as a function of magnetic
field strength}]{The flux-core-temperature relation as a function of magnetic
field strength for a radial field.  The lines follow the solutions for the same field
strengths as in \figref{ts60para}.} 
\label{fig:coretemp0}
\end{figure}}

We have also examined envelopes with conduction perpendicular
to the field for $T_\rmscr{eff}=10^{5.3}$~K, $10^{5.5}$~K and
$10^{5.6}$~K, sufficient to determine the power-law
flux-core-temperature relation for the perpendicular case.
For higher effective temperatures, the
core temperature exceeds $10^9$~K.  For such high core temperatures our
assumption that the core is thermally relaxed breaks down
(\cite{Nomo81}).  For lower effective temperatures, the details of the
equation of state at low densities become important, \ie Coulomb
corrections (\eg \cite{VanR88}).

We fit the flux-core-temperature relation with a power law of the form,
\be
T_{c,7} =  T_{0,7} T_\rmscr{eff,6}^\alpha.
\ee
\tabref{coretemppl} presents the results of the fitting.
\figcom{\bt
\caption{Power-law parameters for the flux-core-temperature relation.}
\label{tab:coretemppl}
\begin{tabular}{c|rr|rr}
B (G) & \multicolumn{1}{c}{$T_{\|,0,7}$} & \multicolumn{1}{c}{$\alpha_\|$} &
\multicolumn{1}{c}{$T_{\perp,0,7}$} & 
\multicolumn{1}{c}{$\alpha_\perp$} \\ \hline
0       & 13.3 & 1.76 &  & \\
$10^{12}$ & 12.0 & 1.76 & 52.9 & 1.21 \\
$10^{13}$ & 10.1 & 1.76 & 123. & 1.16 \\
$10^{14}$ & 9.35 & 1.62 & 213. & 1.17
\end{tabular}
\et}
\comment{
\figcom{\bt
\caption{Power-law parameters for the flux-core-temperature relation.}
\label{tab:coretemppl}
\begin{tabular}{c|rrr|rr}
B (G) & \multicolumn{1}{c}{$T_{\|,0,7}$} & \multicolumn{1}{c}{$\alpha_\|$} &
\multicolumn{1}{c}{RMS Residual} & 
\multicolumn{1}{c}{$T_{\perp,0,7}$} & 
\multicolumn{1}{c}{$\alpha_\perp$} \\ \hline
0       & 13.3 & 1.76 & 0.004 & & \\
$10^{12}$ & 12.0 & 1.76 & 0.018 & 53.5 & 1.21 \\
$10^{13}$ & 10.1 & 1.76 & 0.030 & 123. & 1.16 \\
$10^{14}$ & 9.35 & 1.62 & 0.023 & 214. & 1.17
\end{tabular}
\et}}
The slope of flux-core-temperature relation for the parallel case 
is approximately that found in the analytic treatment of Paper~I
(cf. equations~36 and~37 in Paper~I).  However, the
relationship with magnetic field strength is a much shallower power law
than found earlier.

\subsection{Dipole Fields}

It is straightforward to construct the effective temperature
distribution for a dipole field by interpolating the flux-core-temperature
relation as function of field strength and field inclination.  However,
determining the thermal structure in this manner is less trivial.  The
goal is to determine if an intense magnetic field can cause the envelope
to become oblate through its effect on heat transport, so we will
examine perpendicular and parallel transport for a field strength at the
pole of $2\times 10^{14}$ G by recalculating an envelope solution and
matching the core temperature at the pole and the equator.  For
illustration we choose $T_\rmscr{eff}(\psi=0)=10^{6.4}$~K which yields a
core temperature of $4 \times 10^8$~K.  Along the magnetic equator,
substantially less heat flows through the envelope.  Here,
$T_\rmscr{eff} \approx 10^{5.38}$~K. 
\figcom{\begin{figure}
\plotonesc{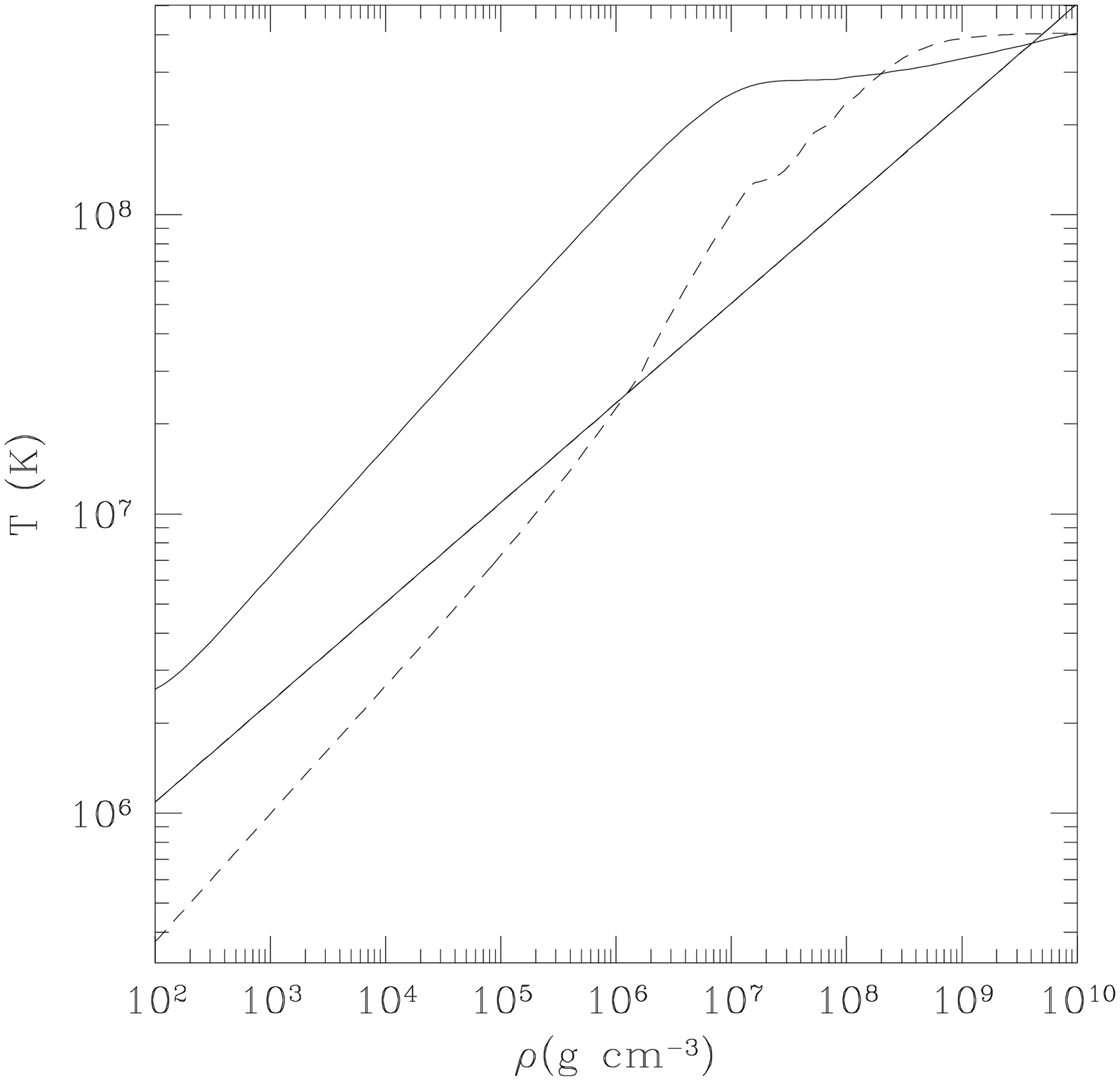}{0.7}
\caption[{The thermal structure for a dipole field configuration}]
{The thermal structure for a dipole field configuration
($B_p=2\times 10^{14}$~G).  The solid
curve traces the solution at the pole, and the dashed curve gives the
results at the equator.  The solid straight line follows the
liquid-solid phase boundary.}
\label{fig:rTdipole}
\end{figure}}

\figref{rTdipole} depicts the two solutions.  The run of temperature
with density is substantially different for the two solutions, so we
would expect that the moment of inertia of the envelope at the pole
would differ from that at the equator.  To first order, the moment of
inertia of the envelope depends on the total mass of the envelope and
the mass weighted mean radius of the envelope.  The envelope is about
0.7 \%\ thinner and less massive at the equator than at the poles.  For a
neutron star with $R=10^6$~km and $I=1.4 \times 10^{45}$~g~cm$^2$, the
moment of inertia of the envelope at the pole differs by a factor of $1.2
\times 10^{-3}$ relative to the equatorial value.   This difference
results in a relative difference in the moment of inertia along and
perpendicular to the magnetic field of $10^{-11}$.

Near the magnetic equator of the neutron star, an equatorial ice band
floats above the ocean.  The ice band extends to a density of $1.6
\times 10^6$~g~cm$^{-3}$ and a depth of approximately 70~cm while the
surrounding ocean reaches $\rho \sim 5 \times 10^9$~g~cm$^{-3}$ and a
depth of nearly 300~m.  This ice band forms because the conductivity in
the nondegenerate magnetized envelope is nearly constant; therefore $T
\propto \rho^{4/9}$ and $\Gamma \propto \rho^{-1/9}$.  The envelope
tends to melt as the density increases.  For any significant ice band
to form the region of the neutron star must insulated from the core by
a tagential magnetic field, \ie near the magnetic equator.  For
free-free opacity without a magnetic field, we obtain $T \propto
\rho^{4/13}$ and $\Gamma \propto \rho^{1/39}$ and this effect is
absent.

\section{Discussion}

The calculations here are patterned after those of \jcite{Hern85}, and
for the case of parallel conduction we reproduce his results with a few
exceptions.  Because we use the formulae of \jcite{Pavl77} to
extrapolate the free-free opacity in the non-degenerate regime beyond
the tabulations of \jcite{Sila80}, we find different thermal structure
in the nondegenerate region than did \jcite{Hern85}, who extrapolated the
calculations of \jcite{Sila80} directly; consequently, for those
especially cool envelopes whose core temperatures depend sensitively on
these opacities, we find that our envelopes transmit less flux for a
given core temperature.   To calculate the thermal conductivity due to
electron transport, we have extrapolated the tabulations of
\jcite{Hern84a} for $n\geq 30$ in manner which maintains continuity
between the magnetized and unmagnetized limits.  Specifically, this
results in a slightly higher electron conductivity for $n \geq 30$ than
\jcite{Hern85} used.  We find that for high effective temperatures,
more flux is transmitted for a given core temperature than
\jcite{Hern85} found.

By examining transport oblique and perpendicular to the field direction,
we have extended the earlier work of \jcite{Hern85} and \jcite{VanR88}
into two-dimensions, and the work of \jcite{Scha90b} to more intense
magnetic fields and more complicated field geometries.  \jcite{Scha90b}
solves the thermal structure equation in two dimensions using the
conductivities of \jcite{Scha88} for $B \leq 10^{11}$~G.  Using the same
set of conductivities, \jcite{Scha90a} treats the cases of parallel and
perpendicular transport using a plane-parallel approximation for $B \leq
10^{13}$~G.   

Rather than solve the two-dimensional thermal structure equations
directly, we have argued in Paper~I that the plane-parallel
approximation holds for the relatively thin envelopes of neutron stars.
The slope of the flux-core-temperature relation for the longitudinal case
agrees with the fit of \jcite{Scha90a} to within 15 \%, and the
normalization agrees to within 8 \%.  Furthermore, \jcite{Scha90a} also
found an upturn in the flux-core-temperature relation for magnetized
envelopes with low effective temperatures.  For the transverse case, we
find that the flux-core-temperature relation has a similar slope as
\jcite{Scha90a} but our models tend to have higher core temperatures.

Our results do not extend beyond an effective temperature of
$10^{6.5}$~K (longitudinal case) and $10^{5.6}$~K (transverse case).
For higher effective temperatures the core temperature exceeds $10^9$~K,
and it is unlikely that the core has relaxed thermally yet
(\cite{Nomo81}). 

\jcite{Scha90b} summarizes the results of the two-dimensional
calculations in terms of fitting functions (\eqref{Schamod}).
In this notation we have examined the applicability of two models.  Both
models have $\alpha=0.5$.  The first fixes $\chi \left ( 90^\circ
\right )=0$, the $\cos^2\psi$ rule, and the second allows $\chi \left (
90^\circ \right )$ to vary, the $a \cos^2\psi + b \sin^2\psi$ rule. 

The two works have only the $T_\rmscr{eff}=10^6$~K with $B=10^{12}$~G
model in common, but our results assume a given flux rule and determine
the change in core temperature.  However, we find that the model
$T_\rmscr{eff}=10^{5.5}$~K and $B=10^{12}$~G is well fit by the
$\cos^2\psi$ rule.  Using the \jcite{Scha90b} interpolation formulae, we
obtain $\alpha=0.44$ and $\chi \left ( 90^\circ \right )=0.32$ which
yields more flux at large values of $\psi$ than we found.  We also
extrapolated the interpolation formulae of \jcite{Scha90b} to
$10^{13}$~G for $T_\rmscr{eff}=10^6$~K and find agreement within 7\%.

Several directions for further work stand out.  The equation of state
at low densities may be affected by Coulomb corrections.
\jcite{VanR88} found that Coulomb corrections play an important role
for effective temperatures less than $6 \times 10^5$~K at $10^{14}$~G.
Their contribution sets in at lower effective temperatures for more
weakly magnetized envelopes.  The prescription of
\jcite{VanR88} for including
Coulomb corrections resulted in negative pressures at low densities.
\jcite{Thor97} have developed a Thomas-Fermi technique which accounts
for the quantization of the electron phase space which may be
applied to envelope calculations without encountering the difficulties
that \jcite{VanR88} found.

In the degenerate regime where electron conduction dominates, the
conductivities are still uncertain.  \jcite{Pote96b} have derived
convenient analytic formulae to calculate the longitudinal transport
coefficients.  \jcite{Pote96b} make slightly different approximations
from other workers.
They include Debye and electron screening in the fluid phase and the
Debye-Waller factor in solid state.  This factor tends to increase the
conductivity over a wide range of temperatures and densities
(\cite{Itoh84}; \cite{Pote96b}).  With an equally complete treatment of the
transverse conductivities, this work could be extended reliably into a
wider range of effective temperatures and magnetic field strengths.

To connect these results with recent observations of isolated neutron
stars (\eg \cite{Grei96}; \cite{Poss96}; and see Table~1 of Paper~I), 
we must calculate how a magnetized atmosphere determines the emergent
spectra from various locations on the neutron star (\cite{Pavl94};
\cite{Pavl96}; \cite{Raja97}), and convolve these spectra with the
effects of gravitational self-lensing to determine the portion of the
neutron star visible as a function of rotational phase (\cite{Page95};
Paper~I).

\section{Conclusion}

We have presented a series of numerical models of neutron star envelopes
calculated in the plane-parallel approximation for $B=10^{12}$ to
$10^{14}$~G, $T_\rmscr{eff}=10^{5.3}$ to $10^{6.5}$~K, and for several
inclination angles of the magnetic field.  We find agreement with
earlier one and two-dimensional calculations, and verify that the flux
along the surface is approximately proportional to the square of the
cosine of the inclination angle for neutron stars with $T_\rmscr{eff}
\sim 10^6$~K.   For hotter and cooler envelopes, this rule provides a
poorer approximation.

With the imminent launch of the AXAF observatory, understanding the
properties of neutron star envelopes is crucial to interpreting the
observations and constraining models of neutron stars, and thereby the
properties of nuclear matter.  The neutron star envelope determines
the thermal flux from the surface at a given core temperature.  The
numerical models presented here will allow a more accurate translation
of the observed emission from neutron star surfaces into knowledge of
their interiors.  When combined with calculations of radiative
transfer through strongly magnetized atmospheres, our models for the
thermal structure of neutron star envelopes will make it possible to
examine the influence of the field strength and surface gravity and
composition on the spectra of isolated cooling neutron stars with
unprecedented detail.

\acknowledgements

We thank the referee, V.A. Urpin, for comments that improved
the presentation.
The work was supported in part by a National Science Foundation
Graduate Research Fellowship, Cal Space grant CS-12-97 and a Lee
A. DuBridge postdoctoral scholarship.

\bibliography{ns,fus,math,physics,qed,mine}
\bibliographystyle{jer}

\figcom{
\end{document}